\documentclass[journal]{IEEEtran}
\usepackage{cite}
\usepackage{amsmath,amssymb,amsfonts}
\usepackage{algorithmic}
\usepackage{graphicx}
\usepackage{textcomp}
\usepackage{xcolor}
\usepackage{booktabs}
\usepackage{caption}
\usepackage{multirow}
\usepackage{array}
\usepackage{balance}
\usepackage{enumitem}
\usepackage{romannum}
\usepackage{longtable}
\usepackage[hyphens]{url}
\usepackage[hidelinks]{hyperref}
\usepackage[utf8]{inputenc}    
\usepackage[T1]{fontenc}    

\setlist[enumerate,1]{label=\arabic*., leftmargin=3em}

\def\BibTeX{{\rm B\kern-.05em{\sc i\kern-.025em b}\kern-.08em
    T\kern-.1667em\lower.7ex\hbox{E}\kern-.125emX}}
    
\begin{document}
\renewcommand{\thetable}{\arabic{table}}
\title{AdaptAuth: Multi-Layered Behavioral and Credential Analysis for a Secure and Adaptive Authentication Framework for Password Security}

\author{\IEEEauthorblockN{Tonmoy Ghosh}\\
\IEEEauthorblockA{Daffodil International University, Bangladesh\\
tonmoy35-932@diu.edu.bd}}

\maketitle

\begin{abstract}
Password security has been compelled to evolve in response to the growing computational capabilities of modern systems. However, this evolution has often resulted in increasingly complex security practices that alienate users, leading to poor compliance and heightened vulnerability. Consequently, individuals remain exposed to attackers through weak or improperly managed passwords, underscoring the urgent need for a comprehensive defense mechanism that effectively addresses password-related risks and threats. In this paper, we propose a multifaceted solution designed to revolutionize password security by integrating diverse attributes such as the Password Dissection Mechanism, Dynamic Password Policy Mechanism, human behavioral patterns, device characteristics, network parameters, geographical context, and other relevant factors. By leveraging learning-based models, our framework constructs detailed user profiles capable of recognizing individuals and preventing nearly all forms of unauthorized access or device possession. The proposed framework enhances the usability–security paradigm by offering stronger protection than existing standards while simultaneously engaging users in the policy-setting process through a novel, adaptive approach.
\end{abstract}


\section{Introduction}
Password security has been an important part of security practice since the necessity of data security was realized. Concurrently, methods for breaking passwords have also evolved in response to the various protection layers and countermeasures developed over time. Despite extensive research and proposals for new authentication paradigms—including password-less methods, whose widespread adoption remains uncertain—the core practice of password-based authentication has seen little fundamental change. On the other hand, traditional password-based authentication is getting riskier day by day, and policies for generating passwords change so often that, to cope with policies, people choose to use easy passwords, which makes them prone to various types of attacks. To avoid risks, awareness for avoiding using weak passwords has risen \cite{Avast2019,Hunt2020,Komanduri2014,Tuerk2019,Hunt2023}, and several industries have started banning common and weak passwords \cite{McDougall2011,Wired2009}, and others are encouraging people to choose strong passwords \cite{Amazon,Google,Microsoft,StopThinkConnect}. Consequently, the adaptive nature of password policies has made password creation cumbersome for users, even as guessing attacks have grown more sophisticated. Moreover, stronger authentication practices like multi-factor authentication have struggled to gain widespread user adoption; even two-factor authentication remains unpopular \cite{Thomson2018}. 

In 2021, a German-born programmer living in San Francisco had ~\$220 million worth of Bitcoin locked in his hard drive, and he could not unlock it because he forgot his password, and there were only 10 guesses allowed to unlock the drive \cite{NYTimes2021,BBC2021}. The specific password remains unknown, making it impossible to determine the proximity of any guess to the correct password. Moreover, the mechanism used to protect that hard drive did not actually disclose what vectors it takes into account to consider an attempt as a potential guess attack. Most authentication servers include some form of limiting mechanism with the aim of safeguarding the user from internet attackers. As a result, the K-strikes mechanism temporarily locks a user account \cite{Pourrahmani2023,Arshi2023,Eyeleko2023} when a repeated incorrect password attempt within the predefined time limit of 24 hours is made. A traditional security-usability trade-off takes place when the K lock parameter is set. Small values of K (such as K = 3) offer stronger protection against online attackers, but they may lead to numerous unintentional lockdowns when an honest user incorrectly inputs (or forgets) their password. The unwanted lockdown rate will be reduced by selecting a higher value of K (such as K = 10), but vulnerabilities to internet attacks may grow. However, while major internet sites are likely to employ some techniques that prevent online password guessing, they do not specify the precise way it is done. Consequently, operators of online services must develop their own proprietary defenses. The impact of password-guessing attacks is so significant that studies \cite{Weir2010,Bonneau2012} have shown that even a small dictionary can be utilized to perform guessing attacks, which can compromise 5\% of accounts. Platforms like GitHub \cite{GitHubBlog}, Twitter \cite{Wired2009}, Apple \cite{Gallagher2014}, and Akamai \cite{Akamai2017} have reported being victimized by this attack, and it remains a security concern that needs more focus. Another technique is still very prevalent in the information security world, which is known by the name CAPTCHA \cite{Kirk2014}, and it was pretty much effective in thwarting automated guessing attacks. However, the advancement of machine learning technology has made the CAPTCHA technique obsolete because recognizing pictures has become very easy for deep learning \cite{Trong2023,Derea2023,Kovacs2023}. The GCHQ Information Security Directorate in the United Kingdom \cite{GCHQ} advises using account lock, throttling, and protection monitoring to defend against automated guessing attacks. Moreover, several other recommendations have been provided by some of the top-level organizations to defend against guess attacks. The Open Web Application Security Project (OWASP) recommended that \cite{OWASP} “All failures are logged and reviewed”, the National Institute of Standards and Technology (NIST) recommended that \cite{Burr2006} “login traffic be monitored for suspicious activity”, The UK’s CESG recommended \cite{GCHQ} “protective monitoring to detect and alert to malicious or abnormal behavior, such as automated attempts to guess or brute-force account passwords” but all the recommended defenses appeared vague and ambiguous and how such recommendations can be adapted to action is unclear at this stage. \\

In addition to the problem discussed above, there is another problem of password security, which is the static password practice. In this practice, passwords exhibit some of the problems discussed earlier, along with the platform-wise policies of password setting. To meet strict policy requirements, people often use password generators to make strong passwords for themselves, which is not a good practice \cite{Adams1999}. Some researchers have claimed that strict password policies do not improve the security of password and there has been some and field studies \cite{Keith2007} and laboratory test \cite{Vu2007} to test that claim and it shows that password restrictions forces users to choose very easy passwords which are very easy to guess and use same password to all over the internet which is very risky because if once get exposed for one account, then all accounts in different platforms will be compromised. These password policies are static and vary based on different platforms' decisions \cite{Hall2023}. This makes the situation more troublesome for users because if a user tries to log in to a platform, then they need to remember what the password was and the password policy if a login attempt fails. A person needs to find out the password policy by creating another account or resetting the password using the “Forgot Password” functionality. Moreover, if the user tries to reset the password, then the password must be different from the previous one, and the user now has to remember \cite{Adams1999} different passwords, which means the problem increases for that user. Additionally, if a user knows the password partially and does not want to reset the password, and tries to log in on that platform, they eventually get blocked with the conviction of performing a guess attack after some failed attempts. We have coined a term addressing all these hassles of password protection, that is: “The Password Dilemma”.\\

\begin{figure}[htbp]
\centering
\includegraphics[width=0.4\textwidth, height=5cm, keepaspectratio]{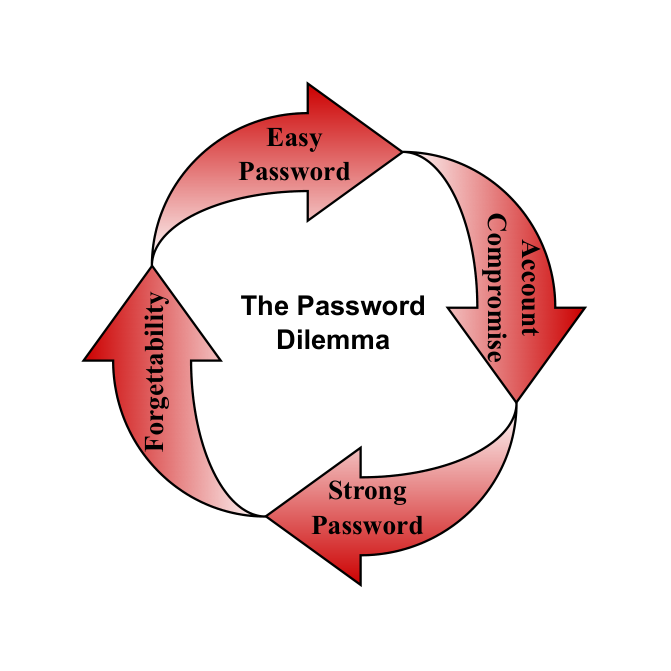}
\caption{Delineating the password dilemma where easy passwords lead to account compromise, account compromise leads to strong passwords, strong password leads to password forgettability, password forgettability leads to easy passwords, and the cycle continues.}
\label{fig1}
\vspace{-2em}
\end{figure}

The problems we have tackled above have been addressed by our two previous papers \cite{Ghosh2025a,Ghosh2025b}, where we tried to provide a solution to both of these problems. In one solution, we talked about password dissection mechanisms where we focus on breaking the password string into several pieces and use them to calculate a user’s legitimacy, and that also helps in preventing getting locked out with a few failed attempts. It provides defense against guessing attacks as well. For another solution, we proposed a mechanism to address the problem of diverse password policies and static passwords, where the passwords will change on the basis of the users’ choice. It could either be on each login or in certain scenarios. In this paper, we merged those two ideas and tweaked them a little bit to provide better performance, and also presented features through analysis on login behavior, personal belongings, geography, and network that aim at developing a complete password security mechanism leading to a centralized authentication mechanism. This will enable the security framework to create user profiles and have a detailed understanding of their password behavior, and provide security with elaborated knowledge, and will be able to identify each individual through their activity with a device. There, any person's behavior profile will be created, and that person's used device's details will be recorded, so our proposed system will be able to identify a criminal wherever they are, no matter how many devices they change. Their behavior with the device will identify them. Similarly, a stolen device can easily be tracked down with this system, as that particular device is profiled with specific data. This system would give users more control over their digital assets and accounts. In addition to that, there are circumstances where we have seen that if a hacker takes control of one’s account and somehow removes the recovery credentials, then there is no way left for the original user to take control back of that account. The current online security practices still rely heavily on this practice, but they cannot do much if scenarios like that occur. There are discussions on Reddit on how to recover a Facebook account if the hacker removes recovery credentials after gaining access to one’s account. This has been a discussion for years, yet Facebook seems not to have taken any action about it. The reporting mechanism about an account getting hacked works in an unknown way. In addition to that, let’s say if a device is profiled with a person’s unique identification, then if that person tries to log in to an account that was previously used by a different profile, that is a suspicious attempt. The main challenge we have to face here is if a legitimate user tries to log in to their account, and if they lose their device to a snatcher, then they will try to log in from a different device. In this scenario, typing-based behavioral analysis becomes critical. With the behavior profile, that person will be identified as a legitimate user, and the system will let them try for an uncertain time. We will explore more applications of this mechanism in human life to make life easier and security stronger. \\
\textbf{Contributions:} The major contributions of this article can be described as follows:
\begin{itemize}
\item Proposed a new method for dissecting password strings that will increase the strength of the dissecting mechanism that was proposed previously.
\item Proposed a new rule to the dynamic password policy mechanism that we named Time Rule.
\item Integrated the Password Dissection Mechanism with the Dynamic Password Policy Mechanism to create a multifaceted security approach.
\item Proposed 173 features, along with some other optional/advanced features, that, with the help of machine learning and artificial intelligence, will lead to building user profiles that will help in identifying legitimate and illegitimate users.
\\
\end{itemize}

The remainder of this paper is organized as follows:- Section II reviews related work on password security, including brute force, dictionary, shoulder surfing, and credential stuffing attacks, along with their countermeasures. Section III introduces the Password Dissection Mechanism, which evaluates password matching percentages and typing irregularities to distinguish benign users from attackers. Section IV presents the Dynamic Password Policy Mechanism, describing the rule-based transformations that adapt password behavior while maintaining usability. Section V details the merged framework, AdaptAuth, integrating both mechanisms and introducing a 173-feature behavioral and credential dataset for decision-making. Section VI discusses the evaluation design and the architecture’s applicability across threat models. Section VII provides the conclusion, Section VIII outlines the limitations of the proposed framework, and Section IX highlights future work directions.

\section{Related Work}
Feature-based password throttling mechanisms \cite{Sandhu2005,Gordon2014} have been proposed, where the features were geographical location, IP address, device information, and others, and these features can be used to feed into machine learning models to help distinguish between malicious and benign login attempts \cite{Freeman2016}. By taking the popularity of bad password guessing into account in decisions to lock down, such defenses as DALock \cite{Blocki2020} and StopGuessing \cite{Schechter2019} provide a better balance between security and usability. It is rare that an honest user who types a password incorrectly will accidentally submit a well-known password. On the other hand, in order to maximize his success rate, an online attacker would like to submit popular password guesses. Therefore, protections like DALock and StopGuessing can immediately lock down an account when an online attacker submits popular guesses frequently, without punishing good users who accidentally write their password incorrectly. A fee-based password verification system has been proposed by Golla et al. \cite{Golla2017}, where a small deposit is required to prove validity, and that fee is refunded if successfully authenticated. Here, they think it is a preventive mechanism because a threat actor would not try to break the password for fear of losing the fee. Schechter et al. \cite{Schechter2010} used a count-min-sketch data structure to find out overly popular passwords and forbid them from being used. Additionally, weak passwords are also forbidden in recommendations of industrial solutions like “Have I been pwned?" \cite{HIBP2019} and “Password CheckUp" \cite{Wired2009}. Florencio et al. \cite{Florencio2007} have found that a 6-digit random PIN can show strong protection against an online attack if the account is locked after 3 unsuccessful attempts for 24 hours. OWASP also suggested account lockout upon proving suspicion to slow down the attacker \cite{OWASP2014}. Bonneau and Preibusch \cite{Bonneau2010} have shown that there is very little practice of account lockout in real life; they did a survey on 150 web applications and found that, among those web applications, 126 permit login even after 100 failed attempts and never locked accounts or IP addresses. Brostoff and Sasse \cite{Brostoff2003} did a survey on 386 undergraduate students and came up with the suggestion of increasing the threshold value for failed attempts from 3 to 10. They also said that setting up a strong password will diminish the risk of allowing 10 attempts. However, this argument has been challenged by empirical studies of analyses of password composition policies \cite{Komanduri2011,Blocki2013}, which show that not all low-entropy password choices are ruled out by many password composition policies. There are others who choose to block IP addresses from where suspicious activity comes \cite{Fail2ban,RFXN,LiquidWeb,TeamPasswordManager}, and in that process, they often block real users. For research purposes, guessing attacks have been done on leaked passwords of millions of people to see how strongly they can bar an attacker from breaking, and in this process, some researchers used algorithms \cite{CMUPGS2015,Kelley2012}, and some used their own built password policy \cite{Weir2010,Weir2009,Weir2010a,Ur2015,Shay2016}. 

In search for password policy guidelines, Vu et al. discovered that the best password-composition policy guidelines for designing passwords are based on small-scale laboratory studies \cite{Vu2007}, while Burr et al. found out that those guidelines also have a basis on theoretical estimates \cite{Burr2006b}. Proctor et al. also found that stricter policies not only make passwords harder to crack but also harder to create and remember, and increasing the minimum length was more effective than applying content constraints \cite{Proctor2002}. From the report of several studies, it can be concluded that if password-composition policies are too demanding (when the policy of generating acceptable passwords is very complex, or when passwords must be changed frequently on a regular basis), users will adopt copying strategies that can reduce both security and productivity \cite{Adams1997,Inglesant2010,Shay2009,Stanton2005}. Several research studies have found that input devices can play a role in password policy and help set a stronger password \cite{Melicher2016,Yang2014,vonZezschwitz2014}. 12-characters and 16-characters password policies have shown the best defense among several other password policies against guess attacks \cite{CMUPGS2015,Kelley2012,Weir2009,Weir2010,Ur2015,Shay2016}. A leaked password-based password policy analysis has been done by Weir et al., where they have proved that entropy-based password policy security measurement is not very helpful \cite{Weir2009}, undermining the study of entropy-based password security measurements \cite{Komanduri2011}. In his study on password policy evolution over a period of 10 years, Steven Furnell found that password policies have not changed that much and are still terrific \cite{Furnell2018}. System administrators found out that people are not properly following password policies after examining leaked passwords, and they also provided a wide range of guidelines for choosing password policies \cite{Florencio2014}. A probabilistic password composition policy has been proposed by J. Blocki et al., who tried to explore human behavior and designed their policy rule on the basis of user input \cite{Blocki2013}.

\subsection{Brute Force Attacks}
Abdelwahab et al. \cite{Abdelwahab2025} worked with multifactor authentication, where on every login, there will be an OTP sent to the client by SMS or email, and then the user will have to enter that OTP, and then that OTP will be hashed and compared with the server’s hash to verify the user. Saputra et. al \cite{Saputra2025}, Reddy \cite{Reddy2024}, and Hamza and Surayh \cite{Hamza2024} proposed to keep the mechanism of setting complex passwords and multi-factor authentication. Reddy \cite{Reddy2024}, as well as Hamza and Surayh \cite{Hamza2024}, also suggested limiting failed login attempts and monitoring login patterns. Hamza and Surayh, in addition to the prior, also suggested implementing learning based IDS, and they provided all their opinions after doing a survey on 22 previous studies. In the proposal of Adamova et al. \cite{Adamova2025}, a central server initializes a global model trained with existing dataset, then this model is sent to client IoT devices, then each device trains the model locally on its own private data, then only the model updates (weights/gradients), not the raw data, are sent back to the server, then the server aggregates these updates to improve the global model, then the improved model is sent back to the devices, and the cycle repeats. Farrel et al. \cite{Farrel2024} leveraged a tool called Wazuh and configured it into their environment to detect brute force attempts and then block suspicious IPs and send an alert through the Telegram app to the authority. Bošnjak et al. \cite{Bosnjak2018} cracked over 99\% of real-world student passwords using various attacks, demonstrating their critical weakness. They recommended ditching weak hash functions like MD5, using strong ones like bCrypt, and adopting the Diceware method to create long, random, yet memorable passphrases. Ruambo et al. \cite{Ruambo2025} designed and tested a Software-Defined Perimeter (SDP) system to stop brute-force attacks on remote access services. They combined Single Packet Authorization (SPA) to hide services, a gateway with deny-all firewall rules and session tracking to only open ports after a valid knock, and Snort IDS to detect suspicious activity. They built a Docker-based prototype, ran brute-force tools like Hydra against it, and showed that their approach significantly reduced CPU load, latency, and packet loss under attack while blocking nearly all brute-force attempts. Adams et al. \cite{Adams2010} analyzed and found the flaws of common protection mechanisms such as Account Lockout ("3 strikes"), IP Address Blocking, and CAPTCHA. They came out working on vectors/directions like Username Direction, Password Direction, IP Address Direction, Knowledge Question Direction, and configured a rate limit mechanism on each of the vectors/directions and imposed temporary blocks for exceeding a certain threshold. Singh et al. \cite{Singh2024} logged login attempt records and made a dictionary of usernames out of it, then blacklist those username and their IPs. Boldyreva et al. \cite{Boldyreva2025} designed a three-party protocol where a client's biometric is strengthened by a helper server using a verifiable oblivious pseudorandom function, and the resulting key material is used to lock a secret into a vault stored on a separate server, with rate-limiting enforced by both servers to prevent brute-force attacks. Jawad et al. \cite{Jawad2025} proposed that after a small number of failed login attempts, then the system will trigger a deception-based mechanism where the attempter will be shown a successful login and will be shown fake data.

\subsection{Dictionary Attacks}
The mechanism of Ashraf et al. \cite{Ashraf2024} proposed SMS encryption with DNA cryptography, and in every session, they send the ciphertext and decryption key to the client. Incident response-based learning proposed by  Huang et al. \cite{Huang2024}, where after an incident, based on the detection systems’ identifying breach value the system educates itself to do better detection in future. Asmat and Qasim \cite{Asmat2019} proposed a mechanism where a user chooses an image and a number to split the image into a matrix bounded by that number, and then the user chooses image chunks from that matrix as login credentials. Kameswara et al. \cite{Rao2018} proposed Spin-Wheel-Based graphical password authentication, where four sub-wheels with numbers 1-36 are under a parent wheel, where the user has to arrange those wheels with chosen numbers. Umejiaku and Sheng \cite{Umejiaku2024} took users' passwords' numerical values to do a Diffie-Hellman-like calculation and then derived a secret key with another mechanism to create a dynamic password to stop brute force attack and dictionary attack. Polpong et al. \cite{Polpong2024} concatenated username and password and numbered them 1 to n and then made username and password of the same length through cycling method and then did modulus calculation and mapped the result with the assigned number and took the value respect to that number then after the operation hashed the final value. Hranický et al. \cite{Hranicky2025} tried to enhance the power of dictionary attack by introducing some rules after clustering similar passwords from existing datasets into groups with the help of machine learning, and then using those rules, they tried to sort out a short list of effective passwords to make password cracking more efficient. Shang et al. \cite{Shang2024} proposed a complex password policy and multi-factor authentication to defend against dictionary attacks. Lin et al. \cite{Lin2025} remained content with the notion that login rate limit and monitoring logins are enough to tackle online dictionary attacks; they focused on offline dictionary attacks in CDN, encrypt the authentication credentials, and use them to validate users in a way that even CDN won’t be able to know about the credentials. Sadat et al. \cite{Sadat2024} supported a complex and lengthy password policy and proposed that users’ names, city, and time to be used after concatenation as passwords.

\subsection{Shoulder Surfing Attacks}
The proposal of Corbett et al. \cite{Corbett2024} needs a Magic Leap 2 headset, ESP32-CAM with a fisheye lens for the rear camera for implementation, and those would be used to detect human eyeball movement to find out if a person is following the device screen. Binitie and Babatunde \cite{Binitie2024} proposed 3 layers of verification that are OTP base verification, then for the PIN code, various sets of random digits will be introduced with actual PIN digits in some of them. The user will select a total of 5 digits from any 2 sets that contain the values. Then the user must answer a security question. Ahmad et al. \cite{Ahmad2025} proposed PassNum, a graphical PIN authentication scheme designed to resist repeated shoulder surfing attacks. They built a 10×10 dynamic digit grid where users authenticate using traversal rules, optional arithmetic operations, color cues, or fake numbers. Through a user study with 32 participants, they evaluated usability (login time, memorability, satisfaction) and security (simulated live and recorded attacks), reporting high accuracy, strong memorability, and complete resistance in their best variation. The study positions PassNum as a potential replacement for conventional PINs by balancing usability and security. Mohamed et al. \cite{Mohamed2024} proposed a mechanism where they customized the display brightness to lower the chance of viewing the inputs from certain angles. Farzand et al. \cite{Farzand2024} systematically reviewed 27 shoulder surfing protection mechanisms, categorized them into ten groups (e.g., icon overlay, haptic, screen brightness), and surveyed 192 UK users to analyze preferences and correlations with personal attributes. They found users valued the mechanisms but leaned toward non-digital alternatives; among digital ones, icon overlay, haptic alerts, and tangible methods were most preferred. Importantly, privacy concerns and tech affinity shaped preferences, while age, gender, and smartphone OS showed no significant impact. Yang and Kong \cite{Yang2024} proposed a graphical PIN protection system where a 3x4 grid is proposed with numbers in those cells, with * and \# included. A grid with those numbers appears on the user's screen, and after seeing the position of the numbers, the user uses a button to hide the grid and draws a line connecting the cells. The user can also draw a line to extra cells to mislead shoulder surfers. In the proposal of Fakheri et al. \cite{Fakheri2024}, for authentication, users have to set a password containing alphanumeric values, then set a sequence of colors from a grid, and then set a sequence of images from a grid. Qin et al. \cite{Qin2025} Image-based graphical credential usage for authentication, where a user has to select multiple images from multiple rounds. During the authentication, the images selected during registration will be shown in 3 rounds along with other images, and even if the user can select a 75\% correct image, it will be considered legitimate. In the proposal of Wu et al. \cite{Wu2024}, Mobile-based touch coordinates, pressed area, pressure, and timestamps data have been captured for right-handers, and some preprocessing algorithms have been used, and then Random Forest and SVM classifiers have been used to make decisions. McConkey et al. \cite{McConkey2024} upgraded an old mechanism for PIN safety from shoulder surfing attacks. There, they changed the two-button functionality in ROTH’s mechanism and changed it to 9 buttons to add more complexity for the attacker.

\subsection{Credential Stuffing}
Pal et al. \cite{Pal2019} worked on a leaked dataset and developed rules to produce similar types of passwords and used them to crack passwords of real accounts to check the validity of their model, and then used that knowledge to suggest to users to choose strong passwords that don’t look similar to their previous passwords. Holthouse et al. \cite{Holthouse2025} proposed rate limiting, complex passwords, reuse prevention, biometric data (e.g., fingerprints or facial recognition), proximity devices (e.g., smart badges), or hardware tokens that generate time-sensitive codes, and two-step verification to prevent credential stuffing. Ajes et al. \cite{Ajes2025} proposed a mechanism effective on the client side for IOT devices where they worked on leaked passwords and developed a Hybrid Similarity Score-based technique that tells users if their passwords are already on the blocklist, if their newly set passwords are too similar to the old one, and if their username and password have similarities. Stejskal et al \cite{Stejskal2024} The author talks about the leaked datasets that can be used to pose severe security threats. Then they discussed some defense techniques to defend against all the attacks, like phishing, password reuse, credentials cracking (brute force), credentials stuffing, exfiltration of software bugs, etc. Their defense proposals are encryption, proper data disposal and archiving, third-party vendor management, employee security awareness, update software and algorithms, develop a cyber breach plan, proper password policies (strong composition rules, account lockout \& mandatory resets, multi-factor authentication (MFA)), and leverage password services (breach monitoring services). Pandey \cite{Pandey2025} talked about the limitations of password-based authentication mechanisms, which are a cognitive burden, and they also stated that user behavior, attack vector proliferation, and organizational cost and complexity are issues. Then the author provided a recommendation of password-less authentication technologies to diminish those problems by implementing biometric authentication, hardware authentication tokens, mobile-based authentication, and certificate-based authentication. Abduhari et al. \cite{Abduhari2025} analyzed several access control mechanisms and did an extensive review of previous literature to find out details about those access control mechanisms. They also did some simulations to find out the effectiveness of those mechanisms, did some quantitative analysis, and came to the decision that multi-factor authentication showed the best performance, followed by role-based access control, and then came strong passwords. Ahmed et al. \cite{Ahmed2025} argued that static credentials and multi-factor authentication have limitations, and they need a dynamic approach. They have taken into account User Behavior Features (Keystroke dynamics, typing speed, and mouse movements), Contextual Features (Device fingerprint, IP reputation, geolocation, login time, and frequency), and Attack-Specific Features (Patterns indicative of brute-force attacks, and credential stuffing) to extract data and apply machine learning to tackle the attempts of account breach dynamically. Pal \cite{Pal2022} proposed Compromised Credential Checking (C3) Services that contain services like Have I Been Pwned (HIBP) and Google Password Checkup (GPC), Frequency-Smoothing Bucketization (FSB), and ID-Based Bucketization (IDB). They also suggested a service like “Might I Get Pwned” (MIGP) – Second-Generation C3 that can detect similar passwords. They also proposed Personalized Password Strength Meters (PPSMs) to find out a particular user’s password similarity compared to the previous one. They also proposed Deployment Practices and Operational Defenses that talk about rate-limiting, breach monitoring and alerting, and blocklisting old and similar passwords. Islam \cite{Islam2025} recommended safeguarding against credential stuffing by leveraging Multi-Factor Authentication (MFA), Breach Monitoring \& Compromised Credential Checking (C3), Rate Limiting \& Bot Detection, Adaptive / Risk-Based Authentication, and Encouraging Strong \& Unique Passwords. Thomas et al. \cite{Thomas2019} tackled credential stuffing attacks by creating a private way to check if your password has been breached. They built a system where your device and a Google server can secretly compare your login info against 4 billion stolen passwords without either one ever seeing the other's data. Their real-world test showed this approach works, with 1 in 4 users who got a warning choosing to create a stronger, new password.

\section{Password Dissection Mechanism}
The Password Dissection Mechanism, as introduced in \cite{Ghosh2025a}, divides the user-provided password string into several segments for comparison with the stored original. Then they took a position number to track every position and its element in a password. Additionally, they calculated the percentage based on the dissected blocks to find out the matched percentage to classify the attempts as either benign or malicious. They also took into consideration keyboard layout to find out possible mistakes that a user could make during login attempts. And, they discussed keeping track of the user's login attempt history to do behavioral analysis. This work builds upon that mechanism by introducing an enhanced password segmentation and analysis technique designed to improve its effectiveness.

\subsection{Dissection and percentage matching}
In the Password Dissection mechanism, the dissection mechanism was arbitrary, and the formal framework that was given to follow could lead to faulty implementation and calculation. That is why we revised it and proposed an improved and stronger framework that will strengthen the dissection mechanism. Previously, the system could choose how many blocks the string was to be divided, such as 1 block, 2 blocks, 3 blocks, and so on, but identifying the mistaken positional values was difficult. In this mechanism, we have brought a method where we will make blocks as well, but in a way where identifying mistaken positional values will be a lot easier, and more security control over the dissection mechanism will be possible to implement. Let’s take a string yomnot2025 that is 10 characters long, and using it, we have shown in the Figure 2 a few ways to dissect that string.

\begin{figure}[htbp]
\centering
\includegraphics[width=0.4\textwidth]{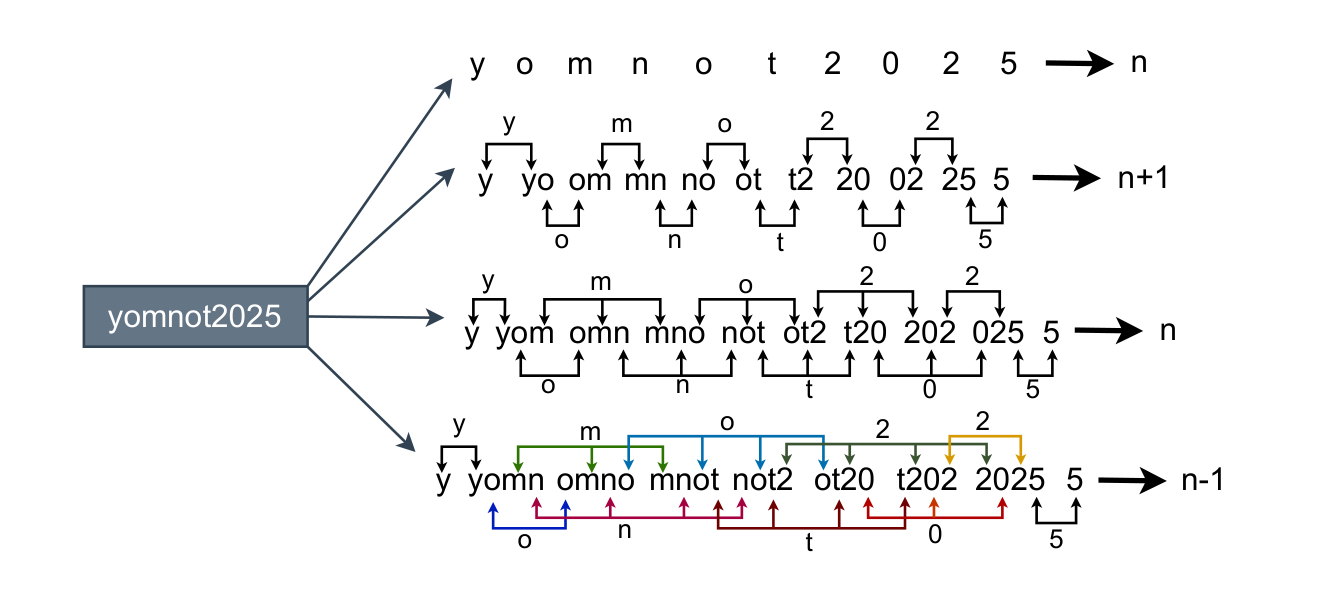}
\caption{Depicting the Reworked Password Dissection Mechanism.}
\label{fig2}
\end{figure}

For any string, we will have a different number of blocks, and because of that, there will be a deterrence on the guessability of the length of the password in case of database compromise. The main focus here is to track every position and the replacement of each value in a password. Therefore, there could be other techniques to dissect a string, and it is the dynamicity of this mechanism. In a password string, if any of the characters are mistyped, then multiple blocks will show errors that could be seen in Figure 2, and from that, the system will calculate exactly where and which values have been mistaken by the user or attacker. Here, we will provide each character percentage based on the number of characters in the string. The percentage will be divided equally among all the characters the password string holds. We can then use these percentages to calculate the matched portions of newly typed passwords with the stored original password. As we said earlier, users might make mistakes because of different policies, and those mistakes might contain 1, 2, or 3 elements for which that user would be considered a guessing attacker in the current implementation of login security posture. This dissection mechanism will handle this problem with the matching percentage; if the percentage is above a certain level, then the user will be considered a benign user. Let’s say, for this, we will consider 80\% or above to be matched to assume the user is benign, and there will be further consideration to make a final decision about a user’s legitimacy.\\
The password dissection and percentage matching mechanism is formally defined as follows:\\
\newline
Let $P_o$ be the original password and $P_u$ be the user's entered password. \\
Let $P_o$ be divided into $N_o$ blocks: $B_{o,1}, B_{o,2}, \ldots, B_{o,N_o}$. \\
Let $L$ be the length of password $P_o$. \\
Let $P_u$ be divided into $N_u$ blocks: $B_{u,1}, B_{u,2}, \ldots, B_{u,N_u}$. \\
\newline
Mismatched blocks: 
\begin{equation}
M_b = \text{match}(B_{u,i}, B_{o,i})
\end{equation}
Pairs of mismatched blocks:
\begin{equation}
M_{b,p} = \{[B_{o,1}, B_{o,2}], [B_{o,2}, B_{o,3}], \ldots, [B_{o,N-1}, B_{o,N}]\}
\end{equation}
Mismatched values:
\begin{equation}
M_v = \sum_{i=1}^{N} M_{b,p,i}
\end{equation}
Match percentage:
\begin{equation}
M_p =\left(\frac{M_v}{P_o}\right) \times 100
\end{equation}
\subsection{Keyboard layout consideration}
Additionally, we incorporated the user's keyboard interaction to find out potential mistakes and understand whether the attempter is desperate or not, whether the attempter is trying all types of values or not.
\begin{figure}[htbp]
\centering
\includegraphics[width=0.4\textwidth]{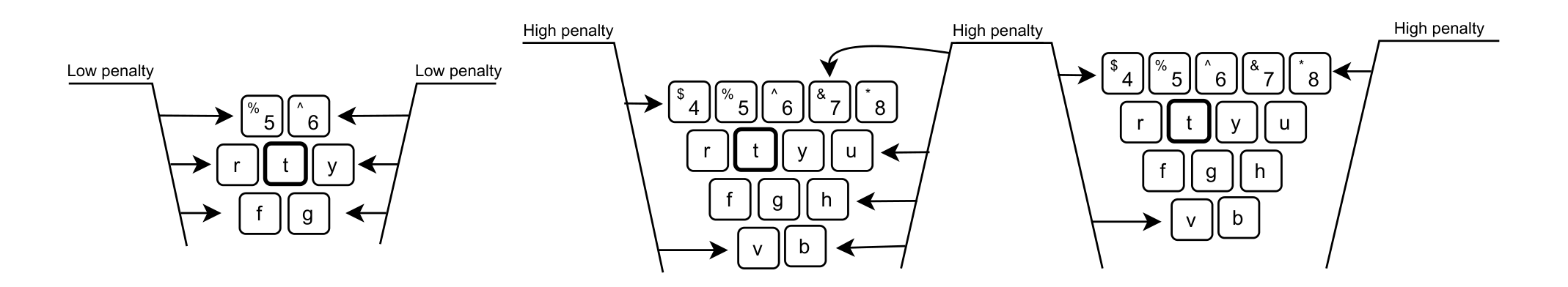}
\caption{Keyboard layout mechanism.}
\label{fig3}
\end{figure}
This data helps us, along with the two mechanisms described above, to separate benign and malicious users. A clear understanding can be obtained with a look at the Figure 3.\\

Continuing from the previous section, if $B_{u,i}$ does not exactly match $B_{o,i}$, but the characters in $B_{u,i}$ are within a predefined keyboard proximity (e.g., adjacent keys) of $B_{o,i}$, a partial match score can be assigned. 
Let $K(v_1, v_2)$ be a function that returns a comparative score (e.g., $0$ to $1$) based on the proximity of keywords between values $v_1$ and $v_2$. 
Therefore, the similarity can be calculated as
\begin{equation}
\text{Similarity} = 1 - 
\frac{\sqrt{\sum_{i} w_i (t_{1i} - t_{2i})^{2}}}
     {\max\left(\lVert T_{1} \rVert_{w}, \lVert T_{2} \rVert_{w}\right)} .
\end{equation}
\subsection{Record of login attempt history}
We tried to record as many events as possible that occurred during failed or successful login attempts. There, we recorded numbers and identity of mismatched blocks, error of new blocks, newly typed value’s distance from previous value, time gaps between login attempts, frequency of character case manipulation, etc. More will be added over time, and that is future work. Here, we worked with both real-time login attempts and historical login attempts, where historical login attempts are the attempts that are recorded from that user’s previous login attempts, with some extra added features. It is like the current login data will be next time’s historical login data. Both the real-time login attempts’ records and the old login attempts’ records are then compared to make a decision about the attempter.

\section{Dynamic Password Policy Mechanism}
In the Dynamic Password Policy Mechanism, we previously proposed a mechanism called the dynamic password policy, where the user's password will be changed after certain events take place. In that dynamic password policy mechanism, several rule-based password policy creation techniques have been introduced, and the dynamic nature of this mechanism comes from those rules. This approach also keeps track of every position of the elements in a password because there will be a capability of altering every position’s value, where a position will be holding user-defined values at different times, and both the user and the system need to know the location. The overall focus of this process is making a simple setup for humans that does not harm usability and security tradeoff. Their rule-based mechanism contains Caesar Cipher rule, Space rule, Leet code rule, Special Character rule, Character Case rule, and Mixed rule. In our opinion, more rules can be added in the future after due deliberation.

\subsection{Caesar Cipher Rule}
In the Caesar Cipher rule, all the Latin alphabets are taken into consideration and each one of them is numbered to have tracking and calculation ability. There are 26 letters, and we have numbered them 1-26 respectively. In Fig. 4, let’s say we have “yomnot” as a password string, and we choose the first position and its value ‘y’ to be manipulated. 
\begin{figure}[htbp]
\centering
\includegraphics[width=0.5\textwidth]{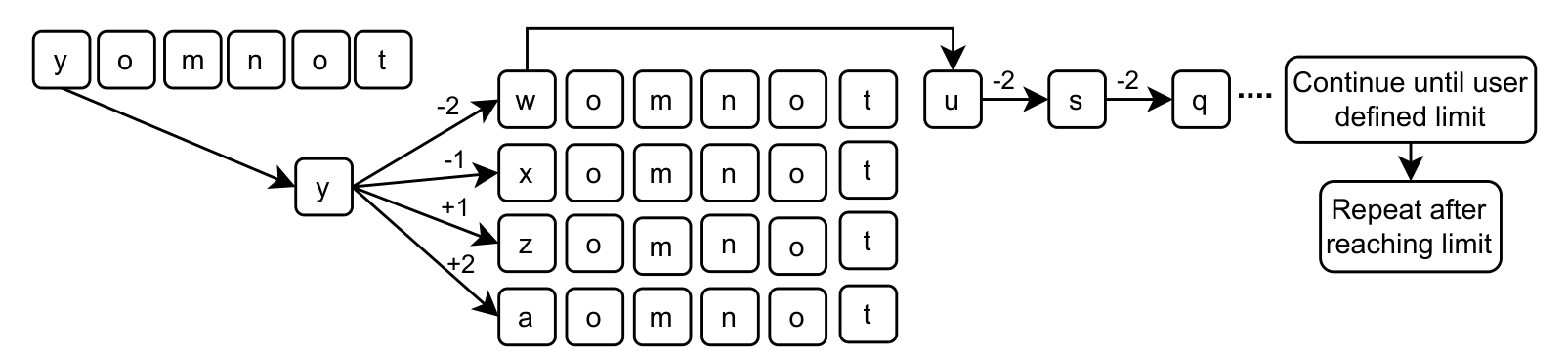}
\caption{Illustration of the Caesar Cipher Rule Mechanism.}
\label{fig4}
\end{figure}
Now we can apply any addition or subtraction to the position value of ‘y’ which is 25 in the Latin script. Here we can find that after adding ‘-2’ to it, we get a new password that will be “womnot” for the next login. And if we keep adding ‘-2’, then we will get ‘u’, then ‘s’, then ‘q’, and it will continue until we reach the user-defined limit. We can also apply the ‘-1’, ‘+1’, ‘+2’, ‘+3’ rule. It totally depends on the user how they want to set the rule. The numeric values 0-9 can also be added to this calculation to have a wide range of value pickup ability. There are 10 numerical values in 0-9, and after adding 10 numerical values to the 26 alphabet values, we will have 36 values in total. Currently, we are not adding special characters because it will complicate the calculations for the users and create pressure on their memorizing ability.

\subsection{Space Rule}
The space rule is about using space in places of values in a password, where on each login, the space can be moved to incremental positions in a password string. This space can be moved in any direction and can be repeated to any position based on the user's selection. In Figure 5, let’s say we have a string “yomnot” and space can stay two times in the first position, then the password will be changed with space in the third position, and then after two times space will be shifted to the fifth position. So, it is totally the user's choice.

\begin{figure}[htbp]
\centering
\includegraphics[width=0.4\textwidth]{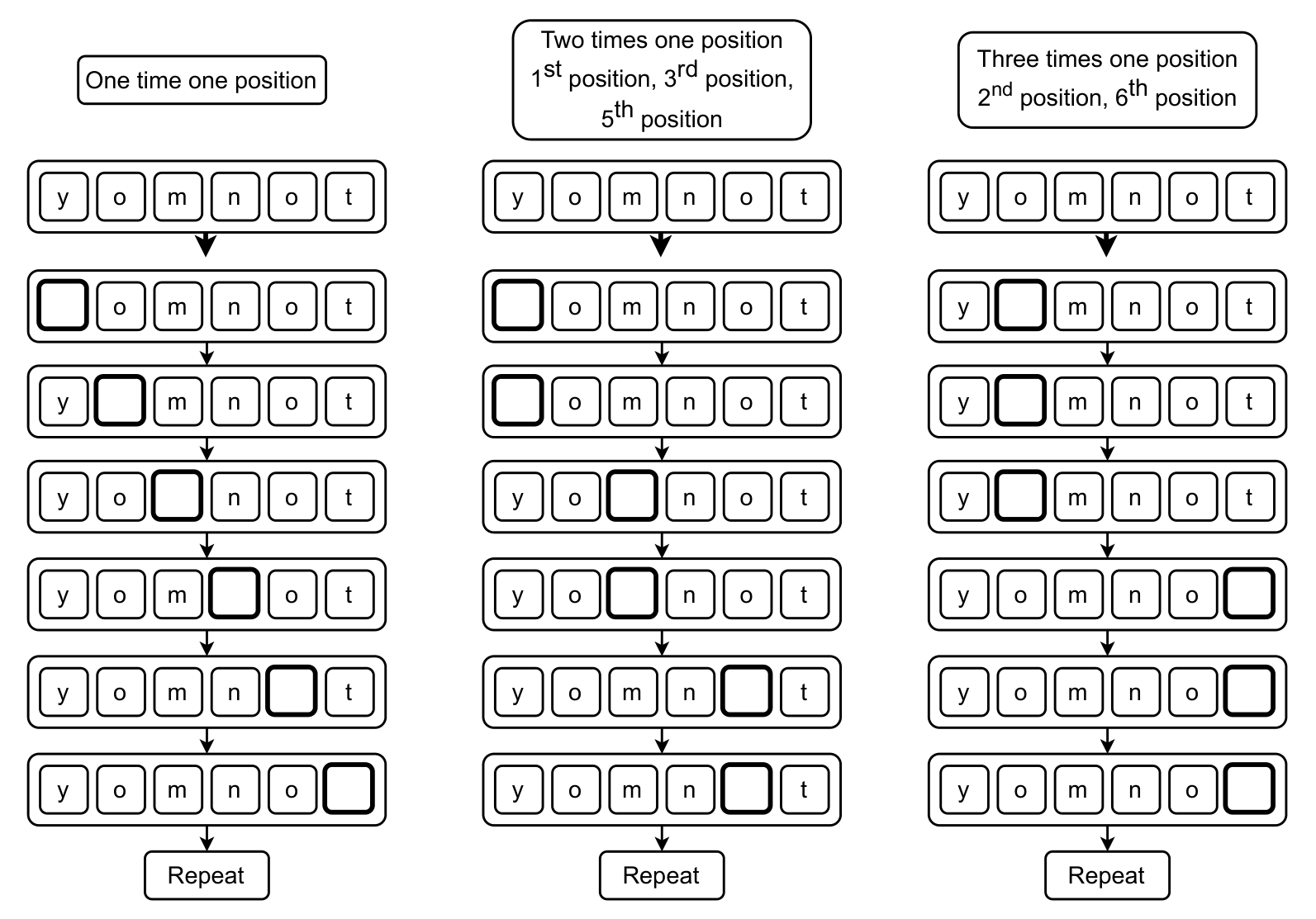}
\caption{Depiction of Space Rule.}
\label{fig5}
\end{figure}

\subsection{Leet Code Rule}
Security professionals or tech geeks sometimes use different types of representation for things they work with. Leet code is that kind of representation of alphabets where they refer to ‘t’ with ‘7’, ‘n’ with '9’, ‘o’ with ‘0’, etc. It is a well-known technique that has been in use for a long time, and one can find it on the internet to enrich knowledge on it. That technique has been incorporated into this Leet code mechanism. Users can also select an alphabet for an alphabet that has a similar pronunciation. In Figure 6, the ‘y’ in the “yomnot” string can be replaced by ‘e’. This shows the flexibility of the user interacting with this technology.

\begin{figure}[htbp]
\centering
\includegraphics[width=0.3\textwidth]{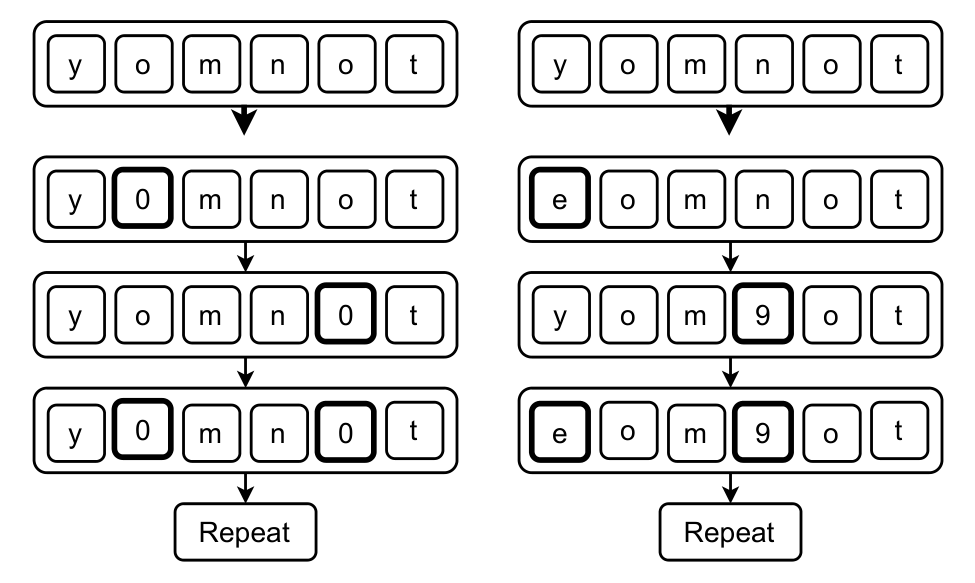}
\caption{Leet Code Rule Mechanism.}
\label{fig6}
\end{figure}

\subsection{Special Character Rule}
In the special character rule, users can use one or a set of special characters to make their password dynamic. The mechanism is more or less similar to what we have discussed so far about the position that one position can have a different value on each login. And multiple positions can also be used for that, based on the user's flexibility. In Figure 7, a random string “yomnot” has been taken to demonstrate the mechanism, where we can see that this rule can be applied with various tactics. On the left side, we can see the position is static, but the values are dynamic, and on the right side, positions and values are all dynamic. It is totally up to the user, based on their convenience.

\begin{figure}[htbp]
\centering
\includegraphics[width=0.3\textwidth]{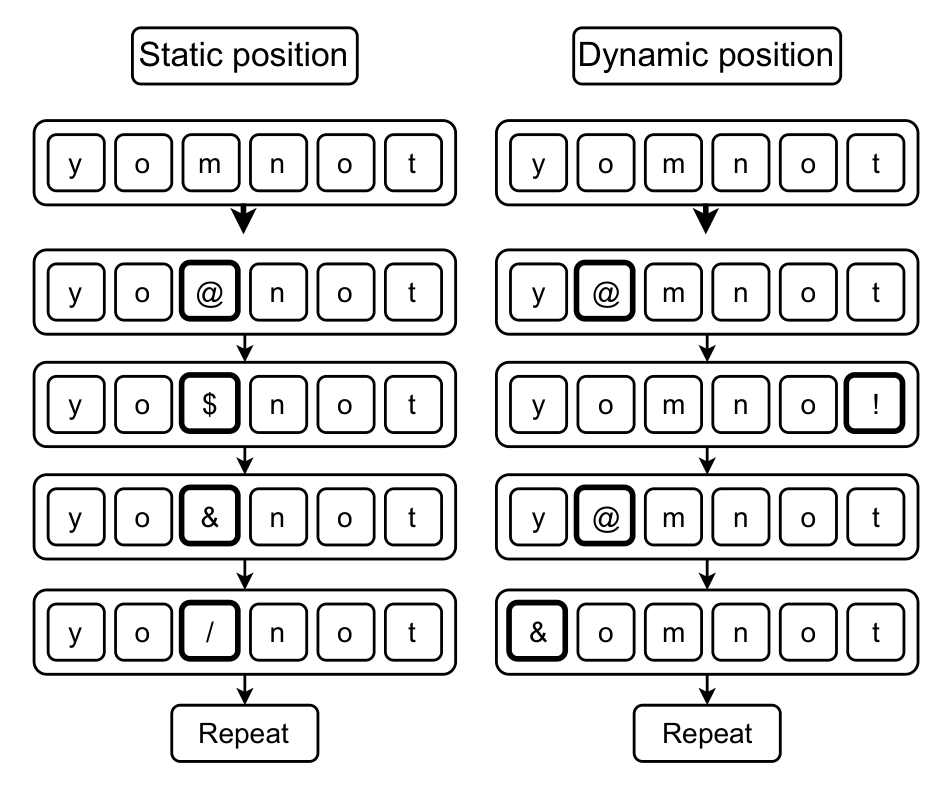}
\caption{Mechanism of special character rule.}
\label{fig7}
\end{figure}

\subsection{Character Case Rule}
In this rule, users can play with the case of the characters in a password string. They will decide on which login, which position’s character should be capitalized, and vice versa. Like in the “yomnot” string shown in Figure 8, after the first login, the password will be “Yomnot”, ‘y’ will be in capital form. Then it could be “yomNot”, “yOmnot”, etc. The user will decide in how many places and how many times this will be applied in a password string.

\begin{figure}[htbp]
\centering
\includegraphics[width=0.3\textwidth]{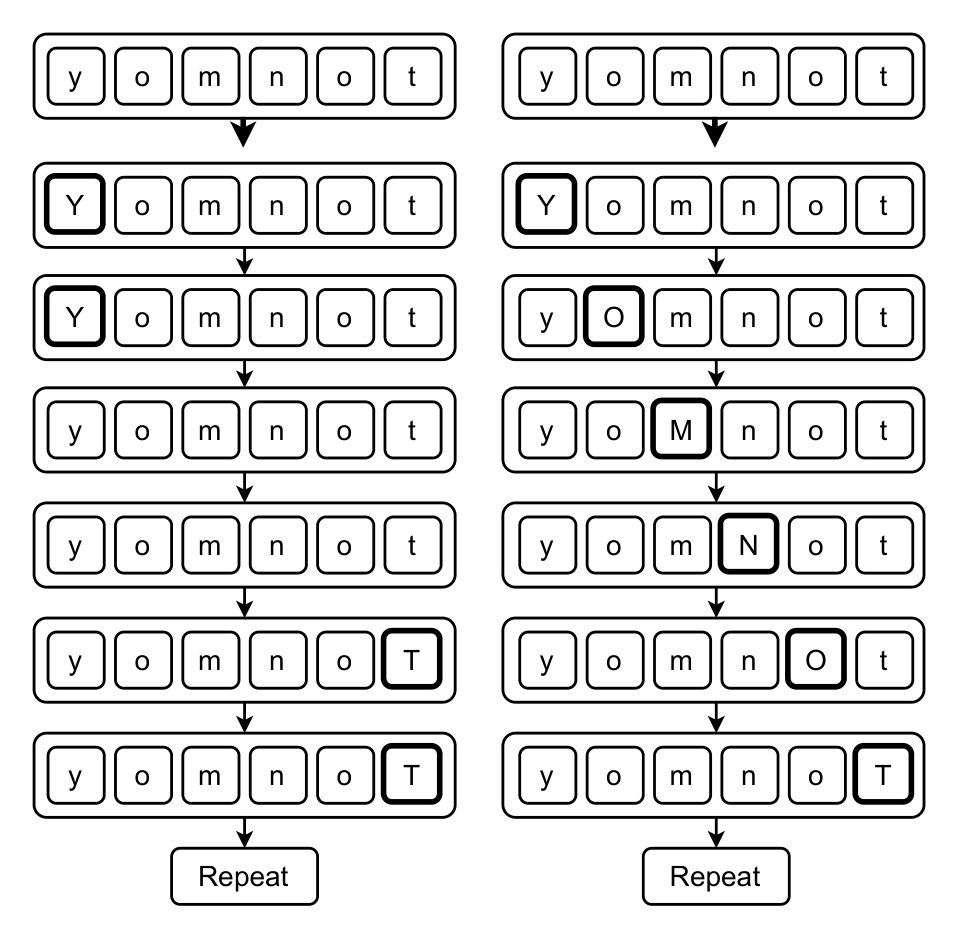}
\caption{Character Case Rule Mechanism.}
\label{fig8}
\end{figure}

\subsection{Mixed Rule}
For mixed rule, the user has the flexibility to use a mixture of all the rules discussed above based on their convenience. The selection of positions also depends on users, as discussed earlier. In Figure 9, we have “yomnot” as a password string, and after applying different rulesets to one password, the positions’ values changed to “1@8Y\#” and “1\&Y9@” based on users’ choices. Here, also like we discussed earlier, positions in the password can be static or dynamic depending on users’ decisions.

\begin{figure}[htbp]
\centering
\includegraphics[width=0.3\textwidth]{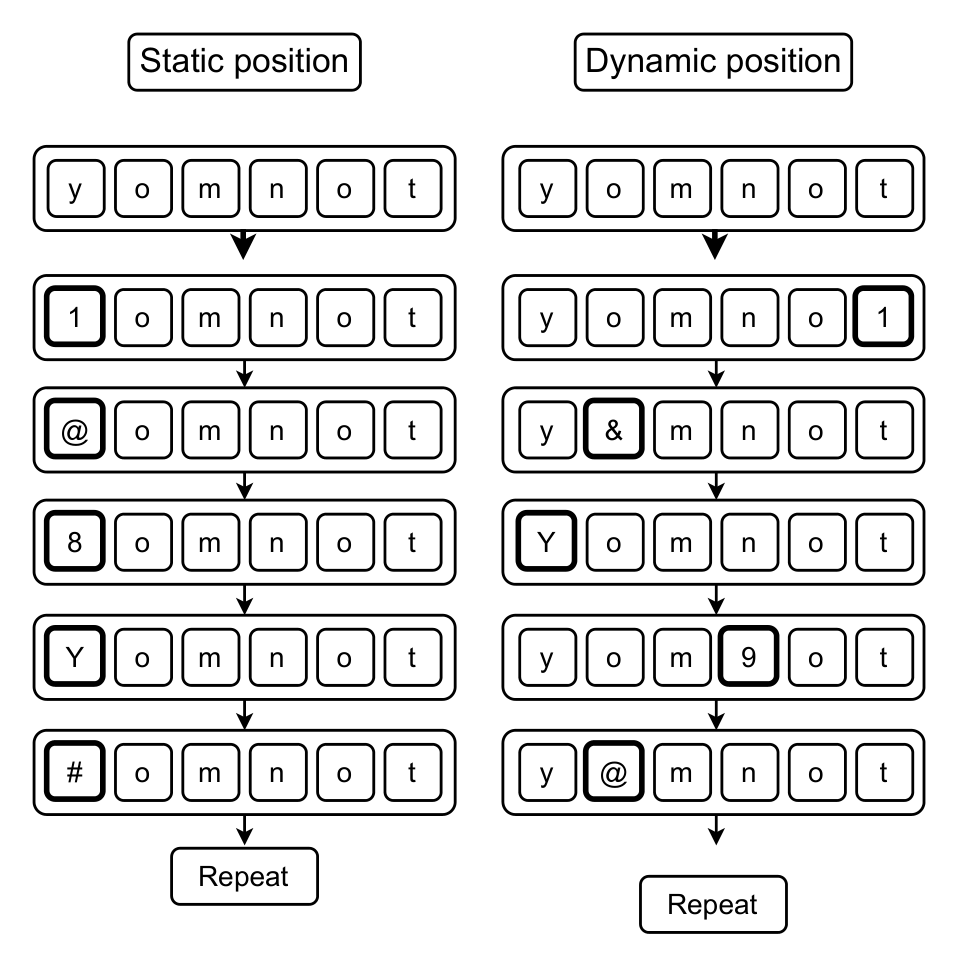}
\caption{Illustration of the Mixed Rule for Setting the Password.}
\label{fig9}
\end{figure}

\subsection{Time Rule}
This is our new addition to this mechanism. In this mechanism, we proposed that the password will be changed on the basis of the time a user initiates the login. That means time will be bound to the login credential where a user can apply addition and subtraction mechanisms to the minute of the real time. Let’s say a user goes for a login at 3:30 PM, and that user has set the rule of adding 15 to the real time, then the login will only be successful if the user enters the value 45 with the main password. If the time is 3:59 PM, then the login value will be 14, which is the next hour’s minute, 4:14 PM. In this mechanism, the main password won’t change; only the logic will change based on the user’s selection. In Figure 10, we have shown a demo for better comprehension.

\begin{figure}[htbp]
\centering
\includegraphics[width=0.4\textwidth]{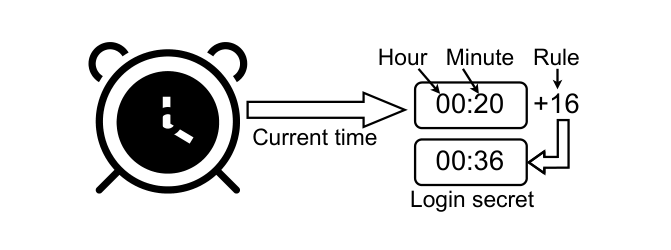}
\caption{Representation of the Time Rule.}
\label{fig10}
\end{figure}
The advantage of this Dynamic Password mechanism is that is scalable, and more rules can be added to that mechanism over time based on user convenience and security approval.\\
\vspace{-0.7em}
\section{Methodology}
\subsection{Merging Password Dissection Mechanism and Dynamic Password Policy Mechanism}
We decided to merge the two ideas and make a complete framework for password security, where we kept intact the goals of those two methodologies and tried to solve the weaknesses they have. The merger of those two methods complements each other’s weaknesses and brings significant improvements to the password protection. The goal of the method discussed in [35] is to bring a solution to the labeling of the attempter as a guess-attacker after a few failed login attempts. And, the goal of the method discussed in [36] is to limit the guessing ability of a malicious user and find out the real user. If a person does credential stuffing and tries to break into another person’s account, then that might evade the guessing attack deterrence mechanism proposed in [35] because the attacker might end up having a decent amount of password matching percentage. In that case, a dynamic password policy mechanism can come in to solve the problem because the attacker will have to try various values to advance further, and that will result in discrepancies with the dynamic rules and the positions where rules have been applied.
\begin{figure}[htbp]
\centering
\includegraphics[width=0.5\textwidth]{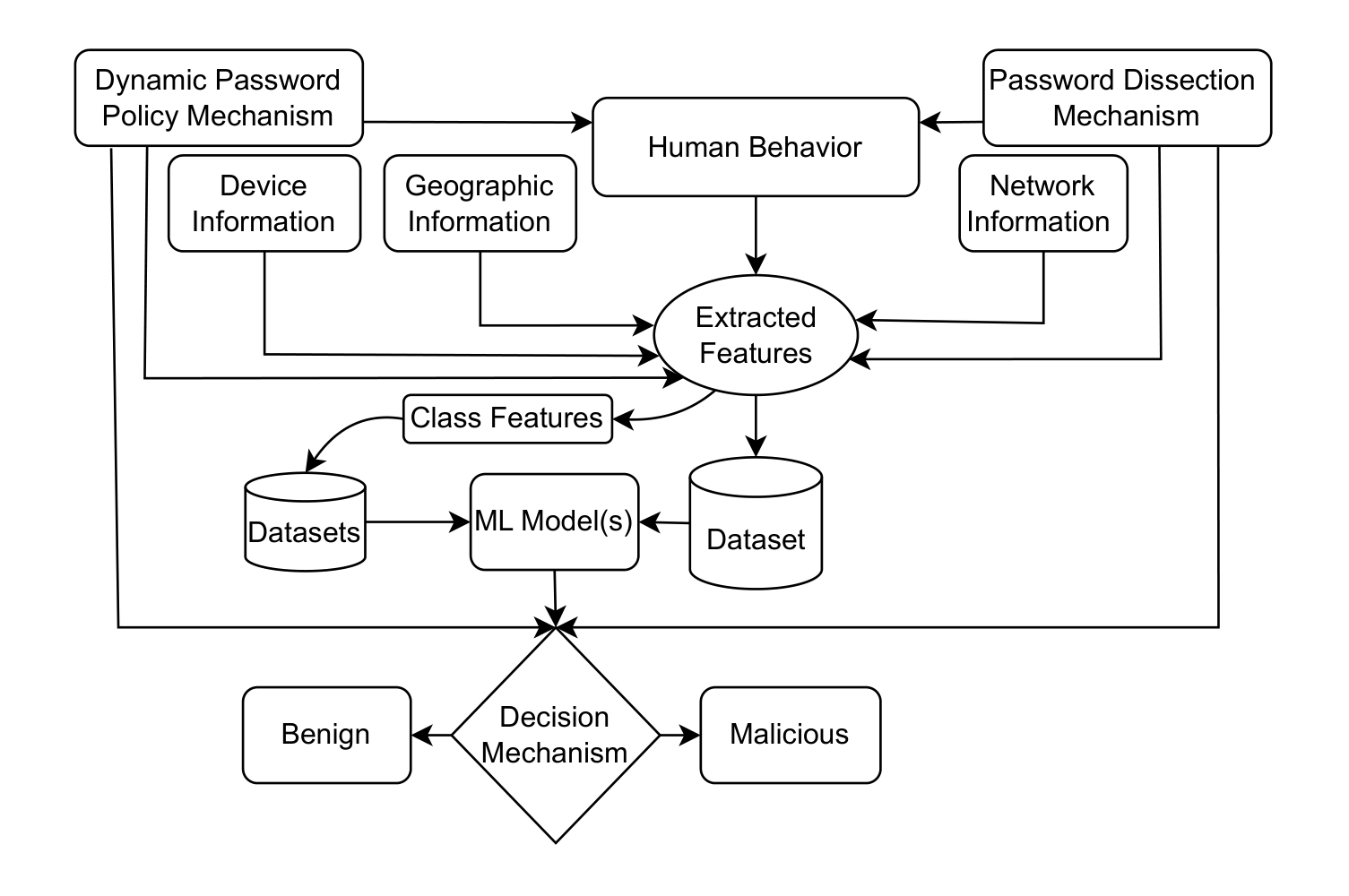}
\caption{Overview of the merging of the password dissection mechanism, the dynamic password policy mechanism, and other decision-making factors' contributions.}
\label{fig11}
\end{figure}
However, if a shoulder surfer observes both the password and the dynamic rule positions, the system might incorrectly authenticate the attacker. In addition to that, due to various rules, we assume that a user will try all the values they have set so far for the dynamic password during login if they forget the pattern, and the platform could still recognize that the user is a guessing attacker or a brute force attacker. Let’s say we have “yomnot2025” as a password string. The special character rule has been applied, and the user has set a set of characters that consists of @, \&, *, and \#. The second position of “yomnot2025” is to be manipulated. After every 4 times of login, the set repeats. If the user forgets this password set, then they could try all the values in the set, and our Robust password security mechanism will consider the user as benign as they matched a certain percentage and got a problem with only one position. Nevertheless, even though the attacker could get very close to breaking the password by posing as a legitimate user, there must be some anomalies to catch that impersonation. We tried to find out those anomalies and proposed 173 features that will record login time activities, abnormalities, irregularities, deviations, and user behavior, and feed them to machine learning models to create user profiles for every single user and their detailed identity and usage. Furthermore, even if an attacker uses a keylogger to discover potential password variations, the feature-based behavioral profile can help detect the impersonation. The system will have minute details about particular users, and it is obvious that a single user’s usage pattern will never match another person's, and this system will capitalize on this inconsistency. This approach also shows significant potential for broader user identity applications, which are explored in the Future Work section.

In Figure 11, we have shown that both the Dynamic Password Policy mechanism and the Password Dissection mechanism have a direct influence on the decision-making mechanism and also have an influence on the features that will come out from the Human Behavior entity. Among all the extracted features, there will be some class features that will hold class data and will have a substantial impact on the decision-making.
\subsection{Features}
\subsubsection{Device Fingerprinting Features}
\begin{enumerate}
\item \textbf{Browser type/version (e.g., Chrome 123.0):} Detects the browser being used (e.g., Chrome 123). A different browser than usual can raise suspicion.
\item \textbf{Operating system and version (e.g., Windows 11, Android 14):} Identifies the OS (e.g., Windows 11, macOS Ventura). A sudden change in OS might indicate account access from a new/unrecognized machine.
\item \textbf{Device type (e.g., desktop, mobile, tablet):} This feature will record the type of device users will use. If a user uses different types of devices, such as a computer, mobile phone, or tablet, then it will be recorded as well.
\item \textbf{Device time:} User spends how much time on which device will be recorded, because this will help in finding out the user's primary device.
\item \textbf{Installed fonts or plugins (where available):} The combination of fonts/plugins provides a quasi-unique signature for the device. Rarely changes unless it’s a new environment.
\item \textbf{Screen resolution and color depth:} These settings are often unique and consistent for a user. If they change, it could signal a new device.
\item \textbf{Touch vs. keyboard input capabilities:} Differentiates between mobile/tablet vs. desktop. An account usually accessed via a desktop, suddenly using touch input, may be unusual.
\item \textbf{User-Agent string:} Combines browser and OS info into one fingerprintable string. Inconsistencies here often flag bots or spoofed browsers.
\item \textbf{Canvas fingerprinting hash (HTML5 feature for subtle device uniqueness):} A rendering-based method that exploits subtle differences in hardware/software to generate a unique hash for a device.
\item \textbf{AudioContext Fingerprinting:} Similar to canvas fingerprinting, but uses the browser's audio stack to create a unique signature.
\item \textbf{Multiple accounts accessed from the same IP in a short time:} If multiple accounts are used to try to gain access to a restricted area, then this raises a red flag about that IP address. And, with the help of the features discussed above about device identification, these types of devices could be identified easily and banned.
\item \textbf{Same fingerprint across many IPs or accounts:} As we said in the previous feature about the bad attempts, if device identification can be achieved, as we discussed in earlier sections, then based on the IP address, we will be able to capture the illegitimate user.
\item \textbf{Missing browser entropy (no screen size, no plugins, etc.):} If there are requests where properties about the browser are missing, then special care should be taken there.
\end{enumerate}

\subsubsection{Geolocation Features}
\begin{enumerate}
\item \textbf{IP-based country, region, city:} Identifies city, region, and country from IP. Large geographical deviations can raise flags.
\item \textbf{ISP/Organization:} The ISP or organization associated with an IPcan be checked to help in the verification of the user's legitimacy. A real user usually logs in from one or two known Internet Service Providers (ISPs) or one organization. An attacker’s ISP/Organization will not be the same as the attacker.
\item \textbf{Latitude and longitude (approximate):} Latitude and longitude data can be used for further identification of a user's geographic location. There is very little possibility that an attempter wanting to log in to another's account will be from the same geographic location.
\item \textbf{Geolocation velocity (distance and time from last known location):} This feature basically tries to find out the comparison between the login time geographic location, where the last successful login time and geographic location compared to the new unsuccessful login time and geographic location of the attempter.
\item \textbf{Region familiarity score (based on previous successful logins):} This feature counts successful logins from locations and tags them as familiar. This number could be helpful in calculating benign and hostile attempts.
\item \textbf{Timezone and system clock offset:} This feature captures the difference between the user's reported system timezone and the expected or historically known timezone associated with their previous logins. It also examines discrepancies between the system clock and the actual current time (system clock offset). Legitimate users typically maintain consistent timezone settings aligned with their location, while attackers may operate from different regions or manipulate system clocks to bypass time-based restrictions. Detecting unusual timezone shifts or clock mismatches can indicate suspicious behavior, helping differentiate legitimate access from potentially fraudulent login attempts.
\end{enumerate}

\subsubsection{Network Attributes}
\begin{enumerate}
\item \textbf{IP address reputation (blacklisted, clean, dynamic/static):} An IP address's reputation—whether it's blacklisted, clean, or dynamic/static—can offer valuable insights into potential adversarial activity. A blacklisted IP often indicates prior involvement in malicious actions such as spamming, botnet traffic, or cyberattacks, suggesting it could belong to a known threat actor. Clean IPs, while not definitive proof of innocence, typically show no history of abuse, making them less suspicious.
\item \textbf{Is VPN detected? (Yes/No):} Detecting VPN usage is crucial in threat hunting and anomaly detection, as adversaries often use VPNs to mask origin, evade geo-blocks, or blend in with benign traffic. However, since legitimate users also use VPNs for privacy, detection must be combined with behavioral or contextual analysis to avoid false positives. In that process, we have to rely on further details that are based on behavioral activity.
\item \textbf{Is Proxy detected? (Yes/No):} Proxy detection can be a strong signal of suspicious activity, especially when correlated with other behavioral anomalies. While proxies have legitimate uses, threat actors often exploit them—particularly public, datacenter, or residential proxies—to mask their true identity, bypass security controls, and automate attacks like credential stuffing or scraping. When proxy usage is combined with unusual login behavior, device fingerprint mismatches, or connections from blacklisted IPs, it significantly increases the likelihood of malicious intent. Therefore, identifying and contextualizing proxy usage is critical for detecting advanced threats and enforcing adaptive security measures.
\item \textbf{Is the TOR exit node? (Yes/No):} Tor exit nodes pose a significant danger when used maliciously because they allow attackers to anonymously route their traffic, making it extremely difficult to trace the origin of cyberattacks. Threat actors often leverage TOR to launch brute-force attacks, distribute malware, exfiltrate data, or communicate with command-and-control servers, all while hiding behind constantly changing, publicly available IP addresses. With the help of public TOR exit node lists, passive network log analysis, JA3/JA3s TLS fingerprint matching, behavioral and heuristic detection, DNS-Based Methods, and detection APIs, this TOR exit node can be detected. All the features we discuss in this paper cover a lot of those analysis techniques.
\item \textbf{Connection type (wired, mobile data, public Wi-Fi):} Connection type can be a strong indicator of suspicious or adversarial activity, especially when it deviates from a user’s normal behavior or originates from high-risk sources. For example, connections from datacenter IPs, VPNs, proxies, or Tor nodes are often used by attackers to mask their identity and location. Unlike residential or mobile IPs typically associated with legitimate users, these anonymous or shared connections are frequently used for credential stuffing, bot activity, or command-and-control communication. Detecting a sudden shift in connection type—such as a user who normally logs in from a residential IP suddenly accessing from a hosting provider—can signal account compromise, automation, or malicious intent.
\item \textbf{ASN (Autonomous System Number) – can help trace institutional access:} ASN (Autonomous System Number) data can provide valuable insights into potential adversary activity by revealing the network ownership and type behind an IP address. If an IP originates from an ASN associated with cloud providers, VPN services, or known malicious infrastructure—rather than a typical residential or mobile ISP—it may indicate anonymized or automated access attempts. For example, logins from ASNs like DigitalOcean or AWS, especially when inconsistent with a user’s normal behavior, can suggest credential abuse, bot activity, or command-and-control operations. By analyzing ASN metadata, defenders can flag high-risk connections, correlate threats across infrastructure, and better understand the attacker’s origin and tactics.
\end{enumerate}

\subsubsection{Temporal Patterns}
\begin{enumerate}
\item \textbf{Time of login (HH:MM)} A user's time of login (HH:MM) is a key behavioral indicator that can help establish a pattern of normal activity. Most users tend to access systems within consistent time windows based on their daily routines or time zones—for example, between 08:00 and 10:00 in the morning. By analyzing and modeling this login timing behavior, security systems can create a baseline for each user. If a login occurs at an unusual time—such as late at night or during hours the user is typically inactive—it may indicate anomalous behavior, potentially pointing to compromised credentials or automated access. This time-based behavior can therefore be used as part of a user’s digital fingerprint to enhance anomaly detection and identity verification.\\
\item \textbf{Day of the week:} Tracking the specific days of the week a user typically logs into a platform can reveal consistent behavioral patterns that aid in detecting anomalies and potential adversary activity. For example, if a user regularly logs in only on weekdays—say Monday through Friday during work hours—a sudden login on a Sunday or holiday may signal suspicious behavior. Such deviations from the user's normal login schedule could indicate unauthorized access, credential compromise, or automated login attempts by an attacker. By monitoring and correlating login days with other contextual signals, security systems can flag unusual activity and strengthen early threat detection. Correlating it with other behaviors will help us find out malicious attempts.
\item \textbf{Mean successful login session starting time window from historical data (e.g., usually between 8–9 PM):} Analyzing the mean login time window from historical data—such as a user typically logging in between 8–9 PM—provides a behavioral baseline that helps identify anomalies. If a login attempt suddenly occurs far outside this usual window, such as at 3 AM, it may indicate suspicious activity, especially if combined with other risk factors like a new device or unusual IP address. Adversaries often operate at odd hours to avoid detection or match their own time zones, making time-based anomalies a strong signal of potential compromise. By comparing current login times with the established average window, security systems can more effectively flag abnormal access patterns.
\item \textbf{Failed login interval variance (compared to normal rhythm):} Collecting the time between the failed login attempts gives us important intelligence that, armed with other details, could lead us to classify between benign and hostile login attempts. There will always be differences between the original user’s attempts and the imposter’s attempts.
\end{enumerate}

\subsubsection{Session/Device History Features}
\begin{enumerate}
\item \textbf{Device/browser familiarity:} If a completely new device is being used against an account to log in, and if the attempt fails, then that is a considerable concern for the system. That information can be used to bolster the decision-making strategy.
\item \textbf{Availability of cookie/token from previous session:} If the Cookie/token from the previous session is available, then we can assume that the user is legit if there is no problem with other data. But if there is any attempt to gain access into the unauthorized area where other data shows inconsistency, then that attempt must be considered a security threat. The chances of occurring session hijacking are the highest because it is one of the safest and easiest ways for an attacker to get access to an account.
\item \textbf{Number of successful logins from current device:} Throughout the whole time since a device was used by a person to access their digital platform, how many times has that device been used to successfully log in to an account? This data is useful in making decisions about whether that device is a user's regular device or not. If a user uses multiple devices, then this data will also help us to determine which device is the primary device and which is most trustworthy.
\item \textbf{First-seen timestamp of device:} The “first-seen timestamp of device” captures the exact date and time a device is initially used to access a user account. This data point serves as a baseline for measuring device familiarity over time. For example, a device that was first seen 8 months ago and has shown consistent usage patterns poses minimal risk, whereas a newly observed device — especially if paired with unfamiliar geolocation or erratic behavior — is treated as high-risk. By monitoring when a device first appeared, the system can detect sudden device usage spikes, correlate with anomalies like VPN usage or abnormal keystroke dynamics, and escalate challenges appropriately. This timestamp also supports forensic analysis by establishing when suspicious access may have begun.
\item \textbf{Changes in system locale or keyboard language settings:} This feature tracks any variation in the system's locale (e.g., language, region format) or the keyboard input language (e.g., from en-US to ru-RU) during a login attempt. These settings are typically stable for most users and align with their geographic and linguistic preferences. Sudden or unexpected changes — such as a Bangla-speaking user’s device switching to a Russian keyboard layout — can signal suspicious behavior, such as account access from a different country, use of compromised systems, or automated attack tools with default configurations. Monitoring these parameters helps differentiate legitimate users from attackers, especially in conjunction with other context-aware signals like IP geolocation or typing behavior. It acts as a low-cost yet effective anomaly detector for impersonation or bot-driven login attempts.
\item \textbf{Login attempt frequency in last X minutes/hours:} Monitoring how often login attempts occur within a defined time window (e.g., 5 minutes or 1 hour) helps distinguish between legitimate and illegitimate behavior. A genuine user typically attempts to log in once or twice, possibly retrying if they mistype their password. In contrast, a high frequency of attempts—such as 10 or more within a few minutes—can indicate brute-force, credential stuffing, or automated guessing attacks. This feature becomes more powerful when analyzed alongside password match percentage, error randomness, and session or network context. Sudden spikes in attempt frequency from unfamiliar devices or networks can trigger additional verification, effectively enhancing the system’s ability to detect and prevent unauthorized access.
\item \textbf{Number of failed logins in a session:} Tracking the number of failed login attempts within a single session provides valuable insight into user behavior and potential threats. A legitimate user might fail once or twice due to a typo or memory lapse, but multiple consecutive failures—especially with varying error patterns—can suggest suspicious activity such as brute-force attacks or unauthorized access attempts. When combined with features like password similarity, typing behavior, and context awareness (e.g., unfamiliar device or IP), a high number of failures becomes a strong indicator of illegitimacy. This feature helps trigger security responses like challenge-based verification, rate limiting, or temporary lockouts to prevent exploitation.
\item \textbf{Login from multiple IPs:} It gives us the details about the user’s usage of the communication medium through which that user can be pinpointed, and we will get an edge if any problem arises with the login.
\item \textbf{Number of logins from multiple IPs, if any:} This record holds the count of the history of the total number of logins from each of the IPs a user uses. With this data, this mechanism will be able to know the primary device(s) of the user.
\item \textbf{Login attempt from unknown IPs:} This feature tells us if there is any login attempt happening from an unknown IP address. Login attempts from unknown IPs tell us about the possibility of unauthorized access or or suspicious activity. This data will be used in correlation with other features to judge the legitimacy. Also, this data will be recorded to compare with any future similar kind of attempt(s).
\item \textbf{Total number of successful logins:} This feature tells us how many times a particular user successfully logged into their account. This gives the system an extra window to take into consideration in the process of decision-making.
\item \textbf{Total number of failed logins:} This feature tells how many times there have been login attempts against an account to break in. It also gives the system a comparative ability in decision-making. This data will also be used in correlation with other data from various features. Based on the number of failed login attempts, the system will have an understanding of the risk against an asset.
\item \textbf{Total number of failed login attempts from known IPs:} This feature will count the total number of failed login attempts from the IPs that are/are familiar. It is usual that users will face failure during login sometimes, and that data will be collected as a failed login attempt from known IPs. This data will also be used after correlating with other data.
\item \textbf{Elapsed time from initiation of login attempt until successful authentication:} This feature will start recording time while the user navigates to the login page. If the user logs in successfully, then this time will be recorded to have a better understanding of the user's memorizing ability and behavior. This data will also be used alongside other data to make decisions by this mechanism.
\end{enumerate}

\subsubsection{Environmental Interaction Patterns}
\begin{enumerate}
\item \textbf{Mouse movement during login:} This feature will be applicable for computer users, where a mouse or a touchpad is mandatory to surf on a page. After navigating to the login page, we would like to track down the pointer movement to gather data about the user's interaction with the mouse or touchpad. This could be class data because we might need information about the speed at which the user navigates to the textbox, and at which speed the user switches to the next textbox after writing the username or email. After clicking on the textbox, the time the user takes to start writing also needs to be recorded.
\item \textbf{Touch data from a mobile device during login:} This is similar to the feature discussed above, but it is for mobile users, as there is no mouse or touchpad used. But the concept is the same.
\item \textbf{Scrolling speed on a particular page:} The scrolling speed of the login page will be recorded because the scrolling speed of each user is different. This data could assist in identifying each individual, and with other data about that user, the identification would be more accurate.
\item \textbf{Window-focus events (e.g., switching tabs before login):} The Window-focus events tells us if the user is switching tabs during login failure.
\item \textbf{Clipboard access detection (pasting passwords vs typing):} This feature will collect data about the user’s password pasting behavior. It is concerning that the user is pasting the password repeatedly, and the login is failing. It also gives us a hint of whether a user is using any kind of dictionary or not.
\item \textbf{Touch event heatmap (for mobile — helps in distinguishing automation/bots):} A touch event heatmap on mobile devices records and visualizes the locations and frequency of user touch interactions across the screen. This data helps in identifying natural human behavior, such as inconsistent finger placements and varying pressure or timing, which differ significantly from the precise, repetitive patterns typically generated by automation scripts or bots. By analyzing these heatmaps, developers can distinguish between genuine user interactions and automated behaviors, enhancing security and improving the accuracy of bot detection systems.
\item \textbf{Click behavior against buttons, textboxes:} This is categorical data that records where in the textbox the user clicks or taps to enable it to take inputs. It is obvious that this behavior varies person to person. Then we can take the intervals between clicking the login button and finishing writing the password. Also, we can take the information about the dwell time of the login button press. In our opinion, this could be effective in classification between original users and fake users. Complements your existing keystroke dynamics and login history analysis.
\item \textbf{Click pattern against a particular page:} This feature records the overall clicks on the overall elements on the login page. A login page contains elements other than textboxes, a submit button that can also be tracked to get a notion about a person’s behavior. All the login pages contain the “Remember Me” option, which allows users to select which login credentials get automatically filled. Moreover, the “Forgot Password” option is also used by users, and it can also be a vector to consider in decision-making. In addition to that, we can also take into account the number of failed tries that prompt a user to go for the “Forgot Password” recovery. In this sense, we can say that this feature holds the potential of being a class data.
\end{enumerate}

\subsubsection{Typing Behavior}
\begin{enumerate}
\item \textbf{Key press and release timings:} Key press and release timings help us to find out the total time needed for a person to complete typping a password string. Moreover, it will also help in finding out the acceleration and deceleration of a user’s typing speed. Further, it could also tell us about the user's digraph-like/trigraph-like behavior. Given the aspects we discussed so far, we can say that the “Key press and release timing” is a class data and will provide us with valuable information.
\item \textbf{Dwell time (duration key is pressed):} Dwell time refers to the duration a key is held down—from the moment it is pressed until it is released—during typing. This metric captures a subtle but consistent aspect of a person's motor behavior, which tends to remain stable across sessions. For example, some users naturally press keys longer, while others type more briskly. By recording dwell times over multiple successful logins, the system builds a behavioral profile unique to the user. During future login attempts, if the dwell time pattern significantly deviates from the established profile, it may indicate an impersonator or bot. Because dwell time is difficult to consciously replicate or spoof, it serves as a reliable biometric feature for distinguishing between legitimate users and attackers.
\item \textbf{Flight time (interval between keys):} Flight time is the interval between releasing one key and pressing the next during typing. It captures the natural rhythm and finger coordination of a user as they transition between keys. Each individual has a unique pattern of flight times based on their typing style, speed, and muscle memory. By measuring flight times across successful login sessions, the system can learn a consistent behavioral signature for each user. During future attempts, significant deviations in flight time—such as irregular gaps or altered typing flow—may suggest that a different person or automated tool is attempting to log in. Since flight time is difficult to imitate accurately, it acts as a strong behavioral biometric for identifying legitimate users and detecting potential intruders.
\item \textbf{Order of positions of mistakes:} The "order of positions of mistakes" refers to the sequence in which incorrect characters appear in a user's failed login attempts. Legitimate users tend to make consistent, patterned errors—often mistyping the same character positions repeatedly due to memory slips, muscle memory, or confusion over password rules. For example, a user may consistently mistype the second and fifth characters of their password in the same order across multiple attempts. By tracking this pattern over time, the system can learn a user's typical mistake order. In contrast, attackers or bots usually generate mistakes at random or in inconsistent positions. Analyzing the order and repetition of mistake positions helps the system distinguish between natural human errors and suspicious, non-human input behavior.
\item \textbf{Typing speed for a full password for every failed login attempt:} With this feature, we aim to extract data that will allow us to determine how a user behaves in terms of typing speed if they fail to log in. This will tell us how the user behaves under pressure of making mistakes or failing.
\item \textbf{Typing speed for a full password for every successful login attempt:} This is similar to the feature discussed just above, but it is for successful logins. This will give the system an understanding of the legitimate user’s flow while typing their password.
\item \textbf{Shift key long pressed or short pressed:} This feature gives us subtle yet important information about the user’s behavior that will help us identify users uniquely. Some users like to long-press Shift keys, and some users like to short-press Shift keys for every word they type. That would be problematic for an illegal attempter to impersonate that behavior.
\item \textbf{Caps Lock button used:} This feature, for computer users, brings information about their behavior with the Caps Lock button. If they use the Caps Lock button, then this aids in building their unique usage pattern, which obviously will be helpful for the system to identify a user through their overall behavior.
\item \textbf{TAB button pressed to switch between textboxes:} If the user uses the Tab button, then this information will be recorded, and this is a rarely used button for a user. Therefore, this information will definitely help the system in identifying legitimate users. This feature is also for computer users.
\item \textbf{Special character and Number switch button:} This feature is for mobile phone users to see if they use that button while writing their passwords. If there is an automation script running to break into one's account, then there is a big chance that this information will not be there, and this can be leveraged to make a distinctive decision.
\end{enumerate}

\subsubsection{Password Characteristics}
\begin{enumerate}
\item \textbf{Length of password during every login button pressed (temporary data):} This feature records the length of the password string while it is fully submitted. Fully submitted refers to the point at which the user has entered the password and clicked the login button, marking the attempt as a final submission.
\item \textbf{Length of password same/bigger/smaller:} This feature records whether the currently entered password is bigger, the same, or smaller than the previously entered password. This is a point to be noted for the system to use in decision-making.
\item \textbf{Number of times password length mismatch:} How many times a user has entered a string value as a password that has mismatched the length of the original password are to be recorded here.
\item \textbf{Number of times length exceeded original password value:} The password comparison we talked about above, if it results in the user-entered password exceeding the length of the stored password, then the occurrence of that event will be counted here.
\item \textbf{Number of times length fell short of the actual password length:} It is the same as the above where we talked about user’s entered string’s length being bigger than that of stored password’s but here in this we will look into the comparative result and count the occurrence of similar event if the user’s entered string has a length that is smaller than the original password’s.
\item \textbf{Incident of appearances of the same length of passwords:} If there are appearances of comparative values where the user-entered password and the stored password are found to have the same length, then it will be recorded here.
\item \textbf{Number of times the same length of passwords appeared:} During all the attempts for login, if there are attempts in which the length of any of the attempted passwords appears to be the same as the stored original password, then the occurrences of those incidents will be counted and stored.
\item \textbf{Positions of mistakes:} If a user fails to log in, then from the next try, this feature will record all the positions where the user will make new entries. This record-taking will continue until the user successfully logs in. Tracking those positions will tell us about the user’s mistake-making pattern, which can be used to identify the benign attempts and harmful attempts.
\item \textbf{Number of positions of mistake(s) in every login attempt:} Places where the user is changing values are an important aspect because they tell the user’s habitual behavior with mistakes in a password string. Therefore, taking mistaken places into account is another tiny but important contribution to the classification mechanism.
\item \textbf{Ambient character or not:} In a session, after failing to log in one time, the system will temporarily keep a record of the password string of failed attempts and afterwards compare the next password string. In those comparisons, for this feature, our system will look for whether the newly entered character is in proximity to the mistaken character that was entered during the very first failed attempt in the current session. This will tell us how desperate the attacker is and will give us an inkling of potential brute force or dictionary attack.
\item \textbf{Character case alteration:} In cases of failed login attempts, the user resubmits the password after modifying the string and then resubmits it for login. This feature takes into account whether the user changed the case of any character in the password string in the process of resubmission.
\item \textbf{Error frequency for a particular position in a session:} This records how many times a position’s value has been changed during the attempts to log in to an account in a session. A particular position’s value change provides data telling a person’s mistakes in that particular position, which could be valuable in decision-making. A user’s common mistakes set a pattern that can easily be recognized and, along with other data, can be used to classify good and bad people.
\item \textbf{Total error frequency for a particular position for all time:} This is the same as the previous feature, where the number of value modification activities in a particular position in a session was said to be recorded. The only extension is that this feature records the adjustment activity that has happened so far since the account creation.
\item \textbf{Number of times character case alteration in a position for all time:} This feature counts the changes in character cases in a particular position of a password string that was submitted by the user during the first login attempt in that session. This data for all the sessions will cumulatively be stored. This tells us how often the user is likely to make mistakes. Also, with this, there might be a possibility of calculating the probability of legitimate users in cases of login failure.
\item \textbf{Number of times character case alteration until a single login button press:} This is almost the same as we discussed in the “number of times character case alteration in a position for all time,” except that this records the count of the modification for every complete password string. Here, the complete password string means that after initiating the writing of the password string, when the user hits the login button, we take that password string as a complete one.
\item \textbf{Number of times character case alteration in a session:} This feature is also similar to the “number of times character case alteration in a position for all time,” except it counts the alteration activity for a session.
\item \textbf{Number of times ambient values are entered in a position for all time:} If a user’s inputted value is ambient to the previously inputted values that led to a failed login, then the count of these types of inputs for all the previous failed login attempts is recorded under this feature.
\item \textbf{Number of times ambient values are entered for all positions combined for all time:} This feature records the summation of all ambient values inputted for all positions of every password string from all the sessions, where all the ambient key presses from every second failed login attempt to the very last failed login attempt in every session will be counted.
\item \textbf{Number of times ambient values are entered until a single login button pressed:} This is almost the same as the feature “number of times ambient values are entered in a position for all time,” except that this records the count of all ambient keys for all the mistaken positions for every complete password string.
\item \textbf{Number of times ambient values entered in a session:} This feature is also similar to the feature “number of times ambient values are entered in a position for all time.” However, this feature sums all ambient key press activities that take place under a session.
\item \textbf{Number of times wrong special character input in a position for all time} With this feature, the data for repeatedly entering wrong special characters in one position in cases of login failures will be recorded. This value will be recorded after the first login attempt fails in all the sessions.
\item \textbf{Number of times wrong special character input for all the positions combined for all time:} A user can have multiple positions in a password string where a dynamic rule can be applied, and that user may need to modify multiple places to try for login. Not only dynamic rules, but also for normal passwords, a user can be confused about the mixture of policies of alphabet, number, uppercase-lowercase, and special characters. This feature will count the sum of the special character mistakes in multiple positions if they are made by the user.
\item \textbf{Number of times wrong special character inputs until a single login button is pressed:} If a user fails to login then multiple login attempts take place and multiple positional inputs of special character could take place for every login attempt. This feature will record the summation of those inputs if this incident takes place. This will also help the system in analyzing the user behavior that ultimately supports user identification.
\item \textbf{Number of times wrong special character input in a session:} Multiple login attempts can take place under a session, and the data obtained from the feature “number of times wrong special character input until a single login button is pressed” will be summed up for all the attempts in a session.
\item \textbf{A user uses single or multi-class values:} After a user sets up a password during account creation, our system examines the password string and determines the number of classes of values the user has included in that password. A password can take different types of inputs, such as characters, uppercase-lowercase characters, numbers, and special characters. For security reasons, we will not record which position has which class. Also, for dynamic password setup, where a position can hold multiple types of values, the positions will not be tracked as well.
\item \textbf{Number of value classes appeared in the current login session(temporary data):} This feature will look for the total number of classes that appeared in the current login session. Mixing it with the user’s password rule and other features could help in distinguishing legitimate and illegitimate traffic.
\item \textbf{Number of positions based on multiple value classes' appearance (temporary data):} When a user tries to log in to an account but fails for the first time in a session, then the system, if there are subsequent login attempts in that particular session, will start to look for changes in classes in all the positions. Changes in character classes in positions could be crucial for scenarios such as credential stuffing attacks and dynamic password-based shoulder surfing attacks. There, if multiple positions have mixed inputs, that will also be known. This feature, and the feature “\textbf{Number of value classes appeared in the current login session}”, could be very important. If we can find a way to permanently store these values, then they will be great contributions.
\item \textbf{Identification of correct values amid heterogeneous inputs at a position:} In cases of repeated login failure, this system will look into the repeated attempts and record if any position that was having a mismatch value, has any correct value appearance while the user is inputting mixed values in that position. 
\item \textbf{Correct input is single or multiple in a position:} If there is an incident like the feature “Does a position have correct input which is having mixed input” occurs, then that correct input is single or multiple in a position will be recorded.
\item \textbf{Multiple positions had correct input:} In all the login failures, if an incident took place where multiple positions had correct input, which was found out during the matchmaking, then that will be recorded here.
\item \textbf{Single correct input is the only one that is entered in the very first:} When a user is trying to log in with multiple attempts, and there is only one correct input during the whole process, and if that is the one entered very first, then that could be a case of Shoulder surfing.
\item \textbf{Single correct input is random:} In a user’s multiple login tries, if there is only one correct input that has come, and if that correct input has randomly occurred in any of the failed tries, then that will be recorded.
\item \textbf{Number of wrong tries before the correct input appears:} When the user keeps trying to log in, if there is one correct input in any position, then that correct input comes after how many failed tries will be recorded here.
\item \textbf{Number of times of having correct input:} How many times correct input appeared among all the failed login tries in a session will be recorded here.
\item \textbf{Number of positions of having correct input:} How many positions having the correct input among all the failed login tries in a session will be recorded here.
\item \textbf{Failed login contains correct password(s) but the sequence is wrong:} If the user tries to log in and the submission contains the correct password, but the sequence does not match, then that will be recorded here.
\item \textbf{Same class character(s)(temporary data):} Do the user's password value changes contain only the same class characters, like the user only trying different alphabets, numbers, or special characters? For a dictionary attacker, there are chances that the same class of values would be used for repeated tries in a position. Along with the distant value and ambient value mechanisms and other features that take input, this feature will help in differentiating legal and illegal users. If a person tries to log in and starts trying to input the same class values, then it might be a triggering point. Using multi-class values could also be a triggering point. User dynamic password rule, and other data with this data could be valuable. If we can find a way to securely store this value, then this will assist in other decision-making as well.
\item \textbf{User input distant value or not:} This is a boolean data type, which checks whether the user enters a distant value or not. When a user fails to log in, then that user tries to do that repeatedly, and in that process, the user modifies values in that password string. If the modified value is very far from the initially inputted value, then this feature will say whether this event has taken place or not.
\item \textbf{Distant value’s distant character(s) entered:} This boolean feature asks whether the user inputted any distant value that is far from a previously inputted distant value. This data also tells us if the user is desperately trying different values to log in, which a valid user would never do.
\item \textbf{Total number of distant value inputs in a position in a session:} This records a user’s total number of distant value inputs in a position in a session through multiple login attempts. Multiple distant values mean the user is not sure about the correct password.
\item \textbf{Total number of distant value inputs in a position for all time:} It is similar to the feature “\textbf{Total number of distant value inputs in a position in a session}” that we discussed above. This feature does some extra work by keeping the record of the sum of the total number of distant value inputs in a position for all time.
\item \textbf{Distant value’s ambient character(s) entered:} We have already discussed what the ambient value is earlier. If a user tries a distant value and fails to log in and then tries that distant value’s ambient value for subsequent logi,n then this feature will pick that up in a true-false mode.
\item \textbf{Number of ambient values of a distant value has been used:} If a user inputs a distant value and then starts inputting the ambient values of that distant value, then this feature will record that. Normally, a valid user would never apply this type of behavior, but there is a high chance that an attacker would behave this way to break password protection.
\item \textbf{Positions where distant values were entered until a single login button is pressed:} The number of positions in which a user who is attempting to log in has provided distant values in a complete password string needs to be recorded to help the mechanism single out that particular user.
\item \textbf{Positions where distant values were entered in a session:} In a session, all the login attempts of a user that contain distant value inputs will be recorded here. In every session, after every first failed login attempt, this data will be collected from the subsequent failed logins.
\item \textbf{Positions where distant values were entered for all time:} After the very first failed login attempt, for every failed login attempt onwards, every distant value input, if there is any, will be recorded. 
\item \textbf{Total position numbers where distant values have been entered until a single login button is pressed:} Previously, we talked about a complete password, where we said that after typing a password, while the user presses the login button, in order to log in, it is considered a complete password. In that process, the positions where the user has entered distant values will be taken into account, and the count of those positions will be recorded with this feature.
\item \textbf{Total position numbers where distant values have been entered in a session:} Same as the feature “\textbf{Total position numbers where distant values have been entered until a single login button is pressed}”, but this records those position-counts for every session.
\item \textbf{Total number of distant value inputs in all positions combined in a session:} How many distant values have been tried by the user during the lifetime of a session will be registered here. Since the very second login attempt, this value will be logged. With this data, we will have an understanding of the extent of a user’s password typing mistakes. Integrated with other feature data, this will help in finding the answer to the question “how long a legitimate user can make mistakes?”
\item \textbf{Total number of distant value inputs in all positions combined for all time:} We can see that this is similar to the feature “\textbf{Total number of distant value inputs in all positions combined in a session}”. It extends its record-collecting ability from one session to all sessions combined.
\item \textbf{Distance level of the tried characters (close, far):} Previously, we talked about ambient character entry in the user’s failed login attempts, where we checked whether the user’s new entry is ambient or not, whether we can take the new failed value as ambient character mistakes of the user, or not. Now the user might try a distant character while trying to log in, and with this feature, we will get to know the spread of the user's key press, and that will help us have an idea about the user’s legitimacy.
\item \textbf{Matching percentage increased/decreased/remained unchanged because of distant value input:} While the user's login attempt contained a distant value, the comparison percentage increased, decreased, or remained unchanged compared to all previous comparisons, and this would be recorded here.
\item \textbf{Keys pressed in a login session (temporary data):} All the keys pressed by the user while typing the password will be recorded temporarily. That will be used to compare values of subsequent login attempts to the previously typed values. This feature will help us more in finding distant values, ambient characters, and in many other ways. We decided not to store these values, as they could pose a threat to the security of the user. Once the database gets compromised, the attacker might get information about other accounts of a user on multiple platforms. However, if we find a way to securely store this information, then the history of this data could play a great role.
\item \textbf{Sequence of key pressing:} After navigation to the login page, the sequence in which a user presses buttons, including the password string, will be recorded. In that, the sequence for the password string will be temporary data because of the security threat we discussed in "\textbf{Keys pressed in a login session (temporary data)}”.
\item \textbf{Password pasting:} This feature will monitor if the user pastes a password in the password field. Pasting a password might raise concerns about adversaries having a password list and taking the password from there or somewhere else.
\item \textbf{Matching percentage increased/decreased/remained unchanged because of ambient key input:} Here, we will look for the state of the password matching scenario after inputting ambient value(s) by a user in cases of failed login attempt(s). If the user inputs an ambient value, then whether the comparison percentage increased/decreased/remains unchanged compared to all the previous password submissions, then this will be recorded. For legitimate users, the percentage should be increased in most cases, remain unchanged in some cases, and decreased in a very small number of cases.
\item \textbf{Ambient value led to login success:} A user might make a mistake in the first password submission in any session, where the inputted value might mistakenly be the ambient value of the actual value. But we take that first input as the original input and subsequent modification to the password string as either an ambient value or a distant value. And whether that ambient input leads to a successful login or not will be registered here.
\item \textbf{Distant value led to login success:} This is almost the same as the previously discussed feature “Ambient value led to login success.” However, a user cannot make multiple mistakes that contain multiple distant values. In those cases, it will be a red flag and could lead to a suspicion of illegitimate users. Incidents can occur where a very few login attempts that contain distant values could lead to login success though it should be a rare case. Even then, if that happens, then it will be registered under this feature.
\end{enumerate}

\subsubsection{Rule Information}
\begin{enumerate}
\item \textbf{Rule name (can be used in security challenges):} This feature records the rule name the user chose to use at the time of setting the password. This is necessary because when that user returns to log in to the account and fails to do so that time, the system will need to know whether the user forgot the password or if there is any intrusion attempt happening. If the attempter’s subsequent attempts go out of the rule that was set by the account’s owner, then that is a problem. Moreover, this can be used in cases of multiple login failures as a security question to proceed.
\item \textbf{User’s frequent mistakes:} We will keep records of the most frequent mistakes that a user makes while typing a password. Usually, a user’s mistakes must be common for that individual, which cannot be matched by others. Also, users’ passwords on different platforms are either similar or very closely related; therefore, mistakes must be common.
\item \textbf{Frequency of rule changes:} If the user switches from one rule to another, then that will be registered here. A User might need to change the old rule setup and set a new rule for obvious reasons. Then this change will be an important information about the user’s behavior and will be very useful.
\item \textbf{Deviated from the rule:} If a user failed to log in, then that user will come under surveillance, where this system will check that user’s previous and subsequent inputs and analyze whether the person deviated from the previously fixed rule or not. This is necessary because a legitimate user will try to stay within one rule, but an illegitimate person will either deviate from the rule or provide inconsistent input. To catch inconsistent inputs, we have discussed mechanisms earlier. Therefore, it is going to be very tough for an illegitimate user to try to break into someone’s account.
\item \textbf{Number of deviations from the rule in one session:} This will record how many times a user has deviated from the rule while attempting to log in under a session. Mixing this with other data will be helpful for the system to make decisions.
\item \textbf{Number of deviations from the rule for all time:} How many times since the very first failed login attempt a user has deviated from the predefined rule will be recorded here.
\item \textbf{Rule repetition threshold (e.g., user rotates rules every 3 logins) (can be used for security question challenge):} In the dynamic setup of a password, values in a password will keep changing after system-defined occurrences take place. A user will set how many times values in a position will change, then repeat that loop, which will be recorded here. And, this data can be used as a security challenge that may arise in a situation to verify the user further. Though our goal is to identify illegitimate and legitimate users with the highest accuracy through our classification model, based on the data that we will collect through surveys.
\item \textbf{Decoy rule existence (can be used for the security challenge):} A user can set a decoy into the password to trick the shoulder surfer and others who want to break the password of that user. This decoy rule will misguide the bad people and help the system filter out these malicious attempters. Using a decoy rule is optional for the users, and if a user sets it, then it will be recorded here.
\item \textbf{Decoy position altered (high red flag):} This is a simple but important test that enhances the ability of the classification models and makes it more accurate if the user integrates this decoy rule into either a dynamic password mechanism or a static password mechanism.
\item \textbf{Position(s) chosen for rule application:} All the positions that have been chosen by the user to apply the rule will be recorded here.
\item \textbf{Position(s) where decoy rule applied:} If a decoy rule has been applied by the user, then in how many positions that decoy rule has been chosen to apply will be recorded here.
\end{enumerate}

\subsubsection{String Dissection}
\begin{enumerate}
\item \textbf{Matching percentage:} During login attempts, the password provided by the user will be broken into chunks and compared with the stored password chunks. The percentage of the matchmaking will be stored to make decisions. Usually, a legitimate user can make one or multiple mistakes in a password string. For that small number of mistakes, we should not call a user illegitimate and impose a block after 5 failed attempts. Moreover, passwords are stored in a database in a Hash format; therefore, we cannot tell how big the user's mistake is. Whether the user is entering a whole different string or just making a mere mistake in the original one is what we have to understand here. And that is why breaking the password into chunks and comparing it to the original chunks will tell us how big the problem is. Using this percentage, we will be able to make decisions that will help the system to more accurately identify good persons and bad persons.
\item \textbf{Position(s) of mismatched values:} After the comparison between the user-entered passwords and the original passwords, the mismatched value positions out of those comparisons will be stored here. These positions are important because a legitimate user would most likely make mistakes in those positions, and an illegitimate user would modify other positions as well, unless that attempter is a shoulder surfer. And if the attacker gathers information through a keylogger, then other features we discussed earlier will assist the system in finding out. Usually, attackers use strings that are either a predefined set of strings or scripts where modification starts from the first position. Also, if there is an attempt to perform a credential stuffing attack, then other features will assist in catching that activity.
\item \textbf{Error increased/decreased/unchanged:} System will look at the matching percentage when the very first attack fails, and then look into the subsequent failed login attempts and log the states of the percentage whether it increased, decreased, or remained unchanged.
\item \textbf{The Percentage of error is unchanged with the new positional problem arising and the old one getting fixed:} It could be a rare case scenario, but it still has the possibility of occurring. This value will also help in profiling a user’s behavior and in identifying desperate login attempts from strangers.
\item \textbf{Position that got fixed:} After the first login attempt failure, in the subsequent attempts, if a position’s mismatch error gets fixed, then that position will be recorded.
\item \textbf{Position that got a new error:} After the first login attempt failure, in the subsequent attempts, if a new positional error appears, then that position will be recorded.
\item \textbf{Number of attempts before solving a positional error:} If a user fails to log in to an account due to a password mismatch, then there could be one or multiple positions where that mismatch occurred. If any individual position gets matched in the following login attempts, then the cumulative number of those attempts will be recorded here. And this will be applicable for every position that was mismatched.
\item \textbf{Number of failed attempts before a successfull login:} When a user fails to log in to an account due to a password mismatch, then that user retries to log in, and if in this process that user gets to log in after, then the number of tries that have taken place before that success will be registered.
\end{enumerate}

\subsubsection{Challenge Pattern}
\begin{enumerate}
\item \textbf{CAPTCHA solving speed based on CAPTCHA type for a single user:} How much time a user takes to solve a CAPTCHA will be recorded here. CAPTCHA is a mechanism that is used by some platforms to distinguish between humans and machines. Though modern techniques are now becoming smart enough to solve basic CAPTCHAs, new types and more complex CAPTCHAs are currently being used by various platforms. This brings more challenges to the users, and they have to be more careful while tackling this challenge. Because of that, CAPTCHA solving speed will vary from user to user, and this will be our takeaway based on each CAPTCHA.
\item \textbf{Average CAPTCHA solving speed based on CAPTCHA type for all users:} The CAPTCHA with the fastest solving rate based on the attempts of all the users will be calculated, and the average speed of solving each CAPTCHA will be recorded here.
\item \textbf{Types of CAPTCHAs a user has tried:} How many types of CAPTCHAs a user has tried will be recorded here. It will help in making decisions on CAPTCHA solving accuracy, percentage, and other calculations.
\item \textbf{CAPTCHA solving accuracy based on individual CAPTCHA for a single user:} For each type of CAPTCHA, how much of it a user can solve properly will be registered here. This will tell us more about the user’s behavior and their capabilities.
\item \textbf{CAPTCHA solving accuracy based on individual CAPTCHA for all users:} For each type of CAPTCHA, how much of it all the users can solve properly will be registered here. This will tell us more about the user’s behavior and their capabilities.
\item \textbf{Session-based CAPTCHA solving accuracy:} The average of the percentages of solving the CAPTCHA will be registered here. The session-based data will help in understanding the user behavior in more detail.
\item \textbf{Overall CAPTCHA solving success rate by a user:} Here, the system will calculate the average CAPTCHA solving success rate by a user. This will provide a quick overview of the engagement of a user with the CAPTCHA.
\item \textbf{Average CAPTCHA solving success rate by all users for an individual CAPTCHA:} This will calculate the percentage of people who faced a particular CAPTCHA and were able to reach an acceptable level of solving that CAPTCHA.
\item \textbf{CAPTCHA complexity classification based on a single user:} CAPTCHA complexity classification can be used to make distinctive decisions about the real and the fake login attempter. Here, we will calculate this complexity by taking into account "\textbf{CAPTCHA solving speed based on CAPTCHA type for a single user}”, and "\textbf{CAPTCHA solving accuracy based on individual CAPTCHA for a single user}”, which will give us a rough idea about the difficulty level of CAPTCHAs from the user’s perspective.
\item \textbf{CAPTCHA complexity classification based on all user:} It is similar to the previous feature "\textbf{CAPTCHA complexity classification based on single user}”, but here we will calculate the complexity by invoking "\textbf{Average CAPTCHA solving speed based on CAPTCHA type for all users}”, "\textbf{CAPTCHA solving accuracy based on individual CAPTCHA for all users}", and "\textbf{Average CAPTCHA solving success rate by all users for an individual CAPTCHA}”. If needed, then more features will be added in the future to calculate the data more accurately.
\item \textbf{Dwell time for CAPTCHA image:} Let’s say a user is given an image CAPTCHA, and that user is trying to solve it. Their taping and releasing of every image in that captcha will be monitored, and the time between those taping and releasing of images will be stored as the dwell time. This will help us in establishing user behavior profiles.
\item \textbf{Flight time for CAPTCHA image:} While solving a CAPTCHA, a user’s time between releasing one image and selecting another will be recorded. This will also help in developing the user's behavior profile.
\item \textbf{Time to solve CAPTCHAs in a session:} When CAPTCHAs appear for the user to solve, the time the user takes to complete the CAPTCHA challenge will be recorded here.
\item \textbf{Time from the appearance of CAPTCHA to start solving:} When a CAPTCHA appears before the user on the device’s screen, then how much time the user takes to start solving it could also be an aspect that can be used in the CAPTCHA-based calculations that we discussed earlier.
\end{enumerate}

\subsubsection{Usage of The Backspace Button}
\begin{enumerate}
\item \textbf{The user uses the backspace button:} This is a Boolean data, and it will check whether any use of the backspace button has taken place or not. When an attacker tries to run a script, it is unlikely they will use the backspace button. And with other features combined with this data, there will be more accurate user-specific identification.
\item \textbf{Number of times the backspace button was used in a complete password:} We previously discussed what we mean by a complete password. In a complete password typing, the number of times the backspace button has been pressed will be counted and recorded under this feature.
\item \textbf{Number of times backspace button used in a session for each user:} In a session, the number of times the backspace button is used by a user will be counted and stored for behavior analysis. It also contributes to the algorithm, which will try to build an understanding that will assist in building legitimate user profiles.
\item \textbf{User empty textbox:} This type of incident can occur while typing a password, where users can make mistakes, and because values change to bullet points as soon as they are entered, the user gets confused about the mistake. Therefore, they might delete the whole typed string instead of one or two last values. This type of trait is very rare for an attacker to perform, and it will be a significant contribution to the user profiling.
\item \textbf{User removed last typed character:} While a user is entering the password, if any wrong value is entered, then the user will remove the last entered value. This activity will be recorded as Boolean data and assist in having a better understanding of the frequent mistakes of users.
\item \textbf{User removed character in the middle:}  If a user removes values from the middle of a typed password string, it will help develop a user profile, and this data will definitely aid in identifying mistakes made by legitimate users..
\item \textbf{Values removed by one backspace button press at a time or long press:} If the user removes all the entered password values, then how does that user remove those values? The user could delete values by a long press or one value at a time. This is a behavioral sign of a person that, with other feature data combined, could help the system notably to find good and bad people.
\item \textbf{Dwell time for the backspace button:} There must be differences in behavior while using the backspace button by different users. Some people delete values slowly and some people do it fast and recording dwell time for pressing the backspace button will facilitate the unique user identification mechanism work more accurately.
\item \textbf{Positions the user used the backspace button:} In a password string in which position’s value has been removed will be recorded here. This will help in building user profiles where this data on using backspace will tell a lot about users’ mistakes. And this data will be aligned with the data that will be collected with other features we discussed earlier about the mistaken positions.
\end{enumerate}

\subsubsection{Complexity Scale}
\begin{enumerate}
\item \textbf{User switch rule:} In this mechanism, there are multiple rules to set a password, and people are free to choose which they want to use. If they don’t find comfort in one rule, then they can shift to another rule. If a user does this, then that activity will be recorded in Boolean format.
\item \textbf{Rule chosen during account creation:} While a user creates an account, the user’s choices of rule will be registered here. This will tell about the popularity and awareness regarding the rules that will help calculate the complexity scale.
\item \textbf{Rule embracement rate for a particular rule:} When people switch to one rule from another, then that data will be collectively recorded here. This data will help the system to calculate the complexity of each rule.
\item \textbf{Rule leaving rate for a particular rule:} Here, people’s choice of leaving a rule will be recorded. Based on the number of people who left any of the rules, the complexity scale calculation, along with other features, could achieve more accuracy.
\item \textbf{Dwell time on a rule that was set during the account creation:} How much time the user stayed with the primarily chosen rule will be logged here. The primarily chosen rule here refers to the feature “Rule chosen during account creation.” This will give a notion about the rules that give a general understanding of the popularity of the rules.
\item \textbf{Time between new rule acceptance and leaving:} There could be a scenario where a user chooses a new rule and leaves the rule after some time. This could happen because people could find a newly accepted rule problematic, then switch to another or return to the previous one. How much time the user stays on a rule will be recorded here.
\item \textbf{Number of times a rule has been chosen by users:} How many times a rule has been chosen by people will be counted here. This data will provide a useful insight into calculating the complexity measurement.
\item \textbf{Number of login attempts failed against every rule, session-wise:} Here, we will count how many times users have failed to log in under a particular rule. And this data will be counted session-wise, which will tell us about the struggles the user had during login.
\item \textbf{Number of login attempts failed against every rule for all time:} The aggregation of all the login failures for each rule will be stored under this feature.
\item \textbf{Return to the previous rule:} If a user switches rules and then comes back to the previous rule, then it will be considered a return. This data will help in the classification of rules in calculating user convenience in rule acceptance.
\item \textbf{Rule false positive:} If a user switches rules and then returns to the old one, then there could be a case where that user stayed for a very short time on a rule and made that comeback. Also, there could be users who, out of curiosity, can change the rules to just have a little bit of experience. Therefore, we have created a separate category to count these types of incidents that will help us while we count “\textbf{Rule embracement rate for a particular rule}”, and “\textbf{Rule leaving rate for a particular rule}”.
\item \textbf{User shifted from dynamic rule to static rule:} If a user shifts from a dynamic rule to a static rule for password, then that data will be collected. This will notify us to further study the user data and try to understand the difficulties that the user was having with dynamic practice.
\end{enumerate}

Above, we have discussed several features that work with the mistakes of users, where they record the mistakes that have been made by the users during previous logins and compare them with the new logins, which helps in finding out discrepancies among the attempts. To have a proper understanding of data, the categorization of data can be done as follows,

Let $E_{\text{new}}$ be the set of positions where mistakes occurred in the current failed attempt. 
Let $E_{\text{hist}}$ be the set of historically common mistake positions for the user. 
The similarity score for mistakes can be calculated using the Jaccard index:

\begin{equation}
J(E_{\text{new}}, E_{\text{hist}}) = 
\frac{\lvert E_{\text{new}} \cap E_{\text{hist}} \rvert}
     {\lvert E_{\text{new}} \cup E_{\text{hist}} \rvert}.
\end{equation}

From this equation, we can find out a user’s mistakes that could contribute to the overall calculation, though in the future, we will try a wide range of possible ways to craft this chunk of user profiling more precisely. Our password dissection mechanism could also be a new contender in the making of this user-mistake-finding-out model.
\\
\subsection{User Signature}
Signature is an advanced class whose contribution could be significant to the user profile construction and security challenges. However, currently, this feature is not available on any device, nor has any software been built to enable us to incorporate this class feature. Although some mobile devices have the feature of pattern lock, our mechanism is not compatible with that feature. Even then, we can consider it as a future technology, and we believe it will have a huge impact on the security posture of individual identity if developed. Signature here will be totally arbitrary based on each user’s perception, and the system should have a detailed understanding and detection ability. A user will be able to generate their own patterns, and this does not have to be the signature that the user uses in documents. Those patterns could also be used as a security challenge because each user’s signature will be different.
\begin{enumerate}
\item \textbf{Signing speed(security challenge):} This will record the time a person needs to sign or make a signature. It should be very difficult for one to match one’s signing time because a sign could have a lot of complex parts that only remain well set in the signer’s brain.
\item \textbf{Deviation occurrence:} This feature will record if any deviation occurred in any attempt.
\item \textbf{Deviation position(s):} This feature will record in which position the deviation occurred. This is an important work where users’ legitimacy and illegitimacy decisions depend heavily.
\item \textbf{Deviation intensity:} This feature is also important because the detection of the intensity of the deviation relies heavily on it. How many mistakes the user has made will be examined here.
\end{enumerate}

\subsection{User Error Detailing}
User error detailing will provide continuous contribution to user profiling, and updates about users’ behavior and uniqueness will be added from time to time. From user error detailing, we will look for some specific answer from our system to check if it is really understanding what we are trying to achieve.
\begin{enumerate}
\item \textbf{User repeating previous positional mistakes:} Under this feature, we will collect data about the user’s mistakes after comparing their past behavior and present behavior. Here we will do some comparative calculations to make this decision.
\item \textbf{User backspace button usage decreases day by day:} Here, the user’s behavioral comparison data will be stored after thorough calculation, where the user’s past data and current data will be compared.
\item \textbf{User backspace button speed increases day by day:} Same as we discussed above, the historical data comparison with current activities to get a rough understanding of the user.
\item \textbf{User mistakes increased/decreased:} Same as the above, we will focus on the past data and present data to have an idea whether the user is improving day by day or having trouble, and whether the relationship is deteriorating or not.
\end{enumerate}

In this mechanism, we will use the security mechanism as a layered approach, which means all the features we have discussed so far will not be triggered at once to find out what is happening. Rather, based on the users’ usage and circumstances, different features will be activated and used to calculate the decision about the situation. Above, we have listed features categorically to have a better understanding of the data, but in reality, a single feature from one category could be called to assist in the decision-making.
\section{Evaluation}
We have talked so far about Password Dissection Mechanism, Dynamic Password Policies, data extraction against user behavior, collecting devices, and other information. We believe that this will make a good password security framework and provide extensive security to the user. Our proposal requires a survey of people to collect data, and it can't be done on artificially created data. This longitudinal study would span several months or years. Participants would require training on the dynamic password mechanism, and their continued engagement would be essential to capturing meaningful behavioral data. This could be done in a university where students will be the participants, and there, their follow-up participation will also be easy. This survey can be done in multiple universities all over the world under a collaboration. Therefore, it needs an arrangement, and it is a matter of sorrow that we currently do not have that kind of arrangement to perform anything like that. A preliminary survey would likely yield insufficient data for conclusive results; therefore, a theoretical evaluation based on established metrics was deemed more appropriate for this stage. However, we have taken a different approach to evaluate our proposal theoretically, where we have reviewed several literatures on various attacks such as Brute force attacks, Dictionary Attacks, Shoulder Surfing attacks, and others. There, we have introduced several metrics to compare those works with our proposed mechanism. Our primary goal is a comprehensive password security mechanism that reduces user burden, improves the user-security relationship, and provides robust security. Therefore, the evaluation metrics are primarily user-centric, that are "User Burden", "User Memorability", "User Involvement", "User Relationship with Security Mechanism", and "System Involvement". We have given a brief explanation below on why we have chosen those metrics.\\
\textbf{User Burden:} Here, we meant that the proposed mechanism could show how much trouble it causes the user. If the password is complex or other means have been introduced to make the security stronger, then that must put pressure on users.\\
\textbf{User Memorability:} This talks about the capability of the user's memorization of passwords. If the password is complex and lengthy, then that obviously decreases the user's memorability of passwords. Users will tend to forget their passwords so soon, and if we take into account the diverse platforms, then it will be more cumbersome for them.\\
\textbf{User Involvement:} This talks about how much the user is involved with the authentication mechanisms. Some techniques rely heavily on the machine doing the security implementation, and the users don't have to do anything. Very little involvement gradually makes people forget about the security postures.\\
\textbf{User Relationship with Security Mechanism:} This talks about the condition of users' relationship with security mechanisms. If the password policy is too complex and lengthy, then the relationship will decrease. If the system does all the work for the authentication, then that will also decrease the relationship because of less involvement.\\
\textbf{System Involvement:} In a mechanism, the involvement of a system in the security implementation has been addressed here. There are a lot of proposals where the operation heavily depends on the system to implement the mechanism.\\

We have also introduced a scale that gives an understanding of the intensity of the metrics we discussed above. In that scale, the measurements are categorized into "Very Low", "Low", "Moderate", "High", and "Very High". Based on the techniques used in early works, we will develop a table on various attacks and compare our work with them there.

\subsection{Brute Force Attack}

\begin{table*}[htbp]
\caption{The Comparative Depiction of Brute Force Attack.}
\label{tab1}
\centering
\begin{tabular}{@{}lccccc@{}}
\toprule
\textbf{Literature} & \textbf{User Burden} & \textbf{User Memorability} & \textbf{User Involvement} & \textbf{User Relationship} & \textbf{System Involvement} \\
\midrule
Abdelwahab et al. \cite{Abdelwahab2025} & High & No involvement & Very Low & Decrease & Moderate \\
Saputra et al. \cite{Saputra2025} & High & Decrease & High & Decrease & Very Low \\
REDDY \cite{Reddy2024} & Very High & Decrease & High & Decrease & High \\
Adamova et al. \cite{Adamova2025} & Very Low & No involvement & No involvement & No involvement & Very High \\
Farrel et al. \cite{Farrel2024} & Moderate & No involvement & No involvement & No involvement & High \\
Bošnjak et al. \cite{Bosnjak2018} & High & Decrease & High & Decrease & Moderate \\
Ruambo et al. \cite{Ruambo2025} & Moderate & No involvement & Very Low & Decrease & Very High \\
Adams et al. \cite{Adams2010} & Moderate & No involvement & Low & Decrease & Moderate \\
Hamza and Surayh \cite{Hamza2024} & Very High & Decrease & Very High & Decrease & High \\
Singh et al. \cite{Singh2024} & Very Low & No involvement & Very Low & No involvement & High \\
Boldyreva et al. \cite{Boldyreva2025} & Very High & No involvement & Very Low & No involvement & Very High \\
Jawad et al. \cite{Jawad2025} & High & No involvement & Very Low & Decrease & Moderate \\
Our Framework & Moderate & Increase & Moderate & Increase & Very High \\
\bottomrule
\end{tabular}
\end{table*}
In the table 1, we have shown some proposals of other authors to tackle brute force attacks and compare them with our work to find out the improvements in the password security practice. Some other works are ingenious and hold good potential to provide better security. However, all of them have some limitations that cannot be avoided, which we have shown in the table with respect to the metrics we have set. Abdelwahab et al. \cite{Abdelwahab2025} said to hash the OTP, and this OTP will be sent to the client on every login by SMS or email, and then the user will have to enter the OTP. This is a problem of dependency, and it increases user burden and affects the relationship between the user and the system, though the user’s memorability and the user’s role have very little effect here. The system has moderate involvement here because it just has to send and hash the OTP to secure the mechanism. Additionally, this OTP sending system also brings the possibility of other attacks on the user account. In their work, Adamova et al. \cite{Adamova2025} proposed a mechanism where they used an existing dataset and learning model to find out brute force attacks on IoT devices. Their scope becomes so narrow for using an existing dataset, which only takes into account a few aspects of brute force attack by looking into network packets. Here, the user burden is very low because the user literally doesn't have to do anything, and the user has no role in this mechanism; that's why no involvement, and the user's memory also has no involvement, but the system has to do extensive work in this mechanism. However, this has a very high involvement in this mechanism. Farrel et al. \cite{Farrel2024} are too dependent on a tool called Wazuh, and based on that tool's decision, they choose to block a user trying to either log in or break in. They did not discuss the situation of blocking real users and took into consideration other aspects as well, which would increase the burden on users. In their mechanism, the system has high involvement, but the user has no extra role beyond traditional password practices. Bošnjak et al. \cite{Bosnjak2018} supported creating lengthy passwords to make them strong, and they also endorsed a mechanism called the Diceware method to create passwords. In our opinion, using that mechanism would force users to write down their passwords in a notepad, and that would increase their burden and affect the memorability of their passwords, though their involvement is high, and decrease the usability-security relationship. Ruambo et al. \cite{Ruambo2025} added an extra layer trying to cloak the services, but increased the user burden in cases of forgetting the password, making several attempts, and getting blocked by the IPS (Intrusion Prevention System). We are not taking the requirement of SPA (Single Packet Authorization) that seriously and as a burden yet, but that could also become a liability, as an attacker can make a fool of it. Also, this system is severely vulnerable to a distributed brute force attack. It doesn't affect the user's memorability, and there is not that much user involvement, but it could decrease the relationship between the user and the security, and in that, most of the work is done by the system. Adams et al. \cite{Adams2010} found issues in the Lockout mechanism, IP address blocking mechanism, and CAPTCHA mechanism. They proposed to log the failure of an individual with respect to username, password, IP, and knowledge question. They attached threshold values to each of them, and upon exceeding those thresholds, the person will get blocked. They did not describe the scenario of a user forgetting the password; they did not tell anything about the policy of setting passwords. Therefore, it could pose a moderate level of burden on the user, and it could decrease the relationship because of the low involvement of the user and some of the complexities we mentioned earlier. Singh et al. \cite{Singh2024} proposed a mechanism where they listed the usernames using what they considered attempts of brute force attack and blacklist them. But they did not discuss if attackers do credential stuffing and try valid username then what they would do. Also, a legitimate user could forget their password and try multiple times to get into the account, then that user could also get blacklisted. And that will certainly put pressure on users, and the relationship with the system will surely decrease. And also, they did not talk about password policy and its implications on users; therefore, we can say that they go with the usual password practice. The technique of Boldyreva et al. \cite{Boldyreva2025} has no involvement of the user's memorability because of the biometric uses, but it poses a serious burden on the user because of its time-consuming computation, and it could decrease the relationship between the user and the system. Also, there are risks of biometric data getting stolen, and once a user's biometric data gets stolen, then that user can never use that feature because a password can be changed, but biometrics cannot be changed. Therefore, in our opinion, biometric data should not be used in security practices. Jawad et al. \cite{Jawad2025} proposed a deception-based mechanism but did not address the chance of legitimate users being recognized as adversaries after a few failed logins. And this will lead to a decrease in usability and a security trade-off. These were so far the unique solutions to the brute force attacks of different authors and their limitations. Now we will talk about the common solution that has been proposed by various authors as well. If we look into the common solution, then we can see that a lot of the authors, such as Saputra et al. \cite{Saputra2025}, Reddy \cite{Reddy2024}, and Hamza and Surayh \cite{Hamza2024} have suggested complex passwords, lengthy passwords, and multi-factor authentication. We have labeled this practice ‘High’ under the User Burden metric because, as we said earlier in the Introduction section, a user has to put extra effort into setting a password with all these things, and the memorability will also be affected because of the combination. When a user’s involvement is high, the relationship with the security practice will also decrease because people could feel apathetic about using it. We also said that authors who added more restrictions to that mechanism, like a login rate limit, will put an extra burden on the user.

\subsection{Dictionary Attack}
\begin{table*}[htbp]
\caption{The Comparative Demonstration of Dictionary Attack.}
\label{tab2}
\centering
\begin{tabular}{@{}lccccc@{}}
\toprule
\textbf{Literature} & \textbf{User Burden} & \textbf{User Memorability} & \textbf{User Involvement} & \textbf{User Relationship} & \textbf{System Involvement} \\
\midrule
Ashraf et al. \cite{Ashraf2024} & Low & No involvement & No involvement & Decrease & Very High \\
Huang et al. \cite{Huang2024} & Very High & No involvement & Very Low & Decrease & Very High \\
Asmat and Qasim \cite{Asmat2019} & High & Increase & High & Increase & Low \\
Kameswara et al. \cite{Rao2018} & Moderate & Increase & High & Increase & Very Low \\
Umejiaku and Sheng \cite{Umejiaku2024} & Low & Decrease & Very Low & Neutral & High \\
Polpong et al. \cite{Polpong2024} & Very Low & No involvement & No involvement & No involvement & High \\
Hranický et al. \cite{Hranicky2025} & No involvement & No involvement & No involvement & No involvement & High \\
Shang et al. \cite{Shang2024} & High & Decrease & High & Decrease & Low \\
Lin et al. \cite{Lin2025} & Moderate & No involvement & Low & Decrease & High \\
Sadat et al. \cite{Sadat2024} & Very Low & Increase & High & Increase & Low \\
Our Framework & Moderate & Increase & Moderate & Increase & Very High \\
\bottomrule
\end{tabular}
\end{table*}

In the comparisn shown in Table 2, Ashraf et al. \cite{Ashraf2024} talked about problems with other encryption mechanisms currently in use and proposed DNA encryption to use. However, DNA encryption itself has its own limitations, and the authors' approach of having a fixed S-box, which, if it gets compromised, then the whole mechanism will crash. Though the authors talked about dynamic per-session tables, there was no indication of how to do that. Also, the authors talked about sending a decryption key to the client, which is not a good idea. In addition to that, there are plenty more issues with this paper, which altogether lead us to the statement that the authors' approach was security by obscurity rather than security through secrecy. In that approach, the users' burden is low because the users don't have to do anything or memorize anything, but the high system calculation and dependency could decrease the users' support for that mechanism. Huang et al. \cite{Huang2024} proposed a mechanism that learns based on the breach incidents and then provides defense in the future against similar events, but they did not take into account the possibilities of an adversarial action being stealthy or the action occurring outside the observation window. Moreover, the incident response-like mechanism seems like a sitting duck that is waiting to get struck and hurt, then keeps vigilance for future similar types of strikes. In this mechanism, the burden on the user is so high that it has to take strikes, and the relationship will definitely be impacted here. Asmat et al. \cite{Asmat2019} and Kameswara et al. \cite{Rao2018} proposed graphical passwords, but in our opinion, those two methods could also put a burden on users and don't diminish the shoulder surfing attack possibility. Also, graphical passwords could get messy when the password needs changing, which could cause the user to forget the sequence. Though here user involvement is pretty good, and those methods would increase the relationship between the user and the security practice. In their proposed mechanism, Umejiaku and Sheng \cite{Umejiaku2024} took the numerical values from passwords and did a Diffie-Hellman-like calculation to do the encryption, but usually people don't use a lot of numbers in a password; they use very few numerical values. And even if they use longer numerical values, then it would be predictable like birthday, birth year, mobile number's portion, etc. Therefore, it is easy for the attacker to break this mechanism and find out the numerical values without much effort, and know the stable string part of this mechanism. Users need to choose lengthy passwords with multiple numbers in them, and this puts a burden on users and damages the relationship. Also, it is prone to a lot of other attacks that the author did not talk about. Polpong et al. \cite{Polpong2024} proposed a username and password concatenation mechanism where they did not take into account the credential stuffing attack, where attackers try to guess usernames along with other credentials. And usernames are not that much of a secret and can easily be obtained through reconnaissance and other means. Also, their used hash functions are not strong and reversible with Rainbow tables. They also did not talk about the length and complexity of user passwords, which might lead to users choosing very simple passwords. We believe that if this system goes into the wild, then there is a very high possibility that an attacker would reverse this mechanism with very little effort. The method that was proposed by Hranický et al. \cite{Hranicky2025} looks like it works on uncertainty and blindly applies sorted passwords that can also be caught if the system is cautious enough. Also, they considered the most common passwords to make their rules and ignored unique passwords. In addition to that, the domain of their mechanism is limited, confined mostly to the English-speaking environment. Shang et al. \cite{Shang2024} and Sadat et al. \cite{Sadat2024} vouched for complex passwords, multifactor authentication, and we have already said earlier what the issues are with these mechanisms. Among them, Sadat et al. \cite{Sadat2024} also talked about concatenating username, city name, and time to the passwords, which makes the authentication more vulnerable. Lin et al. \cite{Lin2025} are certain that login rate limit and monitoring logins are enough to defend against dictionary attacks, along with other attacks, and that's why they focused more on offline dictionary attacks. But rate limit and monitoring are not enough to provide security, and there are plenty of ways to bypass these protections, and those countermeasures could put a burden on users and damage usability to provide more security. Furthermore, their mechanism is ineffective against determined online attackers. If an attacker has a small, high-quality list of candidate passwords for a specific user (e.g., from a password reuse breach), PreAcher's LSH might forward all of them to the origin server.

\subsection{Shoulder Surfing Attack}
\begin{table*}[htbp]
\caption{The Comparative Study Showcases the Shoulder Surfing Attack.}
\label{tab3}
\centering
\begin{tabular}{@{}lccccc@{}}
\toprule
\textbf{Literature} & \textbf{User Burden} & \textbf{User Memorability} & \textbf{User Involvement} & \textbf{User Relationship} & \textbf{System Involvement} \\
\midrule
Corbett et al. \cite{Corbett2024} & High & No involvement & Moderate & Decrease & Very High \\
Binitie and Babatunde \cite{Binitie2024} & High & Increase & High & Increase & Low \\
Ahmad et al. \cite{Ahmad2025} & Moderate & Increase & High & Increase & Low \\
Mohamed et al. \cite{Mohamed2024} & High & No involvement & High & Decrease & Very High \\
Farzand et al. \cite{Farzand2024} & High & Increase & High & Decrease & Very High \\
Yang and Kong \cite{Yang2024} & High & Increase & High & Increase & Moderate \\
Fakheri et al. \cite{Fakheri2024} & Very High & Increase & Very High & Increase & Low \\
Qin et al. \cite{Qin2025} & High & Increase & High & Increase & Moderate \\
Wu et al. \cite{Wu2024} & Very Low & No involvement & Low & No involvement & Very High \\
McConkey et al. \cite{McConkey2024} & High & No involvement & High & Increase & High \\
Our Framework & Moderate & Increase & Moderate & Increase & Very High \\
\bottomrule
\end{tabular}
\end{table*}

From the comparison in Table 3, we can see, In order to protect from shoulder surfing attacks, Corbett et al. \cite{Corbett2024} proposed that users use some external gears that are a huge burden on a user, and obviously decrease the relationship between the user and security posture. Binitie and Babatunde \cite{Binitie2024} proposed 3 layers of defense mechanism where OTP and security questions are common, and the other one, where, in a sequence of random sets, users have to choose their PIN from any 2 sets. It is not very hard for a shoulder surfer to see the PIN because a 5-digit PIN is easy to remember. Moreover, their mechanism is only digit-based, and if they try to expand it to include alphabets and special characters, then it will put an immense burden on users. In the proposed method of Ahmad et al. \cite{Ahmad2025}, they provided several techniques to counter shoulder surfing, and among them, in our opinion, is strong, and that is the Arithmetic and Traversal rule, which will increase users' memorability, involvement, and relationship with security. For other techniques in this mechanism, either cells are fixed or PIN is fixed in the grid, and this system depends heavily on the digits. Though those techniques also would increase involvement, memorability, and relationship, their limitations discard them. Mohamed et al. \cite{Mohamed2024} proposed to customize display brightness to limit visibility from angle to prevent shoulder surfing, but in different environmental scenarios and for different types of users, the performance of their mechanism raises a big concern about the usability. And, we believe, the complexity will definitely affect the user's affinity to the technology. Farzand et al. \cite{Farzand2024} did some digging in the previous works on shoulder surfing and tried to do a short survey on some UK people who said that they prefer a non-digital prevention system. This implies that they are not satisfied with current implementations, which clearly tells that they see current implementations as a burden and their relationship with security is not very good. Some of the people chose Icon overlay, Haptic alerts, Tangible mechanism, and Screen brightness adjustments. However, we can say that self-report cannot be the same as the behavior, and preference cannot prove effectiveness. And those techniques people preferred are not in security practice, and also have their own limitations. Yang and Kong \cite{Yang2024} proposed a mechanism where a user has to draw lines connecting cells that contain the correct PIN digits. First of all, their mechanism is confined to digits, ‘*’, and ‘\#’, and that limits this mechanism’s strength. Secondly, though the numbers’ positions are shuffled every time the user tries to enter the PIN, there is still a possibility that the observer gets an idea about the password. Moreover, the decoy technique they talked about could lead unauthenticated persons to get authenticated accidentally. Also, hiding the grid could increase the chance of making mistakes for a user, thus increasing the burden. Though through this mechanism the interaction between user and system increases, the memorability and involvement of users increase, but these are not enough to compensate for the other major flaws. Fakheri et al. \cite{Fakheri2024} proposed a 3-layered shoulder surfing prevention system that only increases burden on users, where they incorporated traditional password practice, then color sequence, and then image identification. It will damage the user's memorability and usability-security relationship, though it has good user-security interaction traits. Qin et al. \cite{Qin2025} proposed a graphical authentication system where images appear in multiple rounds and users have to select the correct images to get authenticated. There are also chances of making selections of images messy from time to time. Also, due to the round system, there are chances that users might choose images that are easy to remember, and a shoulder surfer can also see which images that person is selecting. Though this system could increase user memorability, involvement, and relationship, it could lead users into problems. Wu et al. \cite{Wu2024} did operations on mobile devices where they considered several metrics to collect data and use machine learning to find out user patterns. The limiting factors here are multiple, such as their mechanism is only confined to mobile devices, and thus cannot be widely accepted. In addition to that, the domain of their analysis of user behavior is very narrow, and they did not take a lot of other vectors into consideration, and ignored several aspects of threats and vulnerabilities. Their data is also limited to right-handers, and that adds more to the shortcomings of this methodology. McConkey et al. \cite{McConkey2024} polished an older mechanism and modified it to increase the effectiveness of the defense against shoulder-surfing attacks. However, there is a high chance of users getting tangled while trying to input passwords during login. The authors introduced 9 black buttons and stated that users can use multiple buttons to deceive the shoulder surfer, but they did not really test how many of those buttons put how much burden on users. They did do a test in a confined environment with a few participants, but that did not cover enough. Even if it could increase user involvement, it would severely damage the usability.

\subsection{Credential Stuffing}
\begin{table*}[htbp]
\caption{The Comparative Study Table of Credential Stuffing.}
\label{tab4}
\centering
\begin{tabular}{@{}lccccc@{}}
\toprule
\textbf{Literature} & \textbf{User Burden} & \textbf{User Memorability} & \textbf{User Involvement} & \textbf{User Relationship} & \textbf{System Involvement} \\
\midrule
Pal et al. \cite{Pal2019} & High & Decrease & High & Increase & Moderate \\
Holthouse et al. \cite{Holthouse2025} & Very High & Decrease & Very High & Decrease & Moderate \\
Ajes et al. \cite{Ajes2025} & High & Decrease & Moderate & Decrease & Moderate \\
Stejskal et al. \cite{Stejskal2024} & Very High & Increase & High & Decrease & High \\
Pandey et al. \cite{Pandey2025} & High & No involvement & Very Low & Decrease & High \\
Abduhari et al. \cite{Abduhari2025} & High & Decrease & Moderate & Decrease & Low \\
Ahmed et al. \cite{Ahmed2025} & Very Low & No involvement & No involvement & Decrease & High \\
Pal \cite{Pal2022} & High & Decrease & Low & Decrease & High \\
Islam \cite{Islam2025} & Very High & Decrease & High & Decrease & High \\
Thomas et al. \cite{Thomas2019} & High & Decrease & Low & Decrease & High \\
Our Framework & Moderate & Increase & Moderate & Increase & Very High \\
\bottomrule
\end{tabular}
\end{table*}

From the comparison shown in Table 4, we can say that, Pal et al. \cite{Pal2019} and Ajes et al. \cite{Ajes2025} looked into older leaked datasets and sorted out some passwords, and using that knowledge, they built a tool that made people aware of weak passwords. Their approach does a good job of admonishing people about the danger. However, their approach could also lead people to set complex passwords that they did not talk about, and it could put a burden on users, although their mechanism, in our opinion, would increase user-security relationships. Moreover, like we said earlier, their work heavily depends on the most common passwords of the same category; thus, they missed out on a lot of unique passwords that could also play a good role in breaking passwords. Furthermore, since their work is based on leaked datasets of a few organizations, we can say that their model is more inclined towards the English-speaking masses. Holthouse et al. \cite{Holthouse2025} also endorsed rate limiting, complex passwords, reuse prevention, biometric data, hardware tokens, and two-step verification. And we discussed earlier that all of these techniques put a very high burden on users, and biometric data could be stolen, and lost biometric data is most problematic. Users' memorability gets impacted, user-security relationships get impacted, but user involvement gets increased. Stejskal et al. \cite{Stejskal2024} gave a total guideline of how a company should organize to tackle cyber attacks, including password-based attacks, but did not talk about any innovative technology. Rather they stayed dependent on current technologies and thinking encryption, proper data disposal and archiving, third-Party vendor management, employee security awareness, update software and algorithms, develop a cyber breach plan, proper password policies (strong composition rules, account lockout \& mandatory resets, multi-factor authentication (MFA)), leverage password services (breach monitoring services) are enough to stand strong. These are current technologies that exist in practice; even then, defending against password-based attacks is still relevant. Also, they did not analyze which of their proposals could cause people to put a lot of effort or labor into keeping up with the standards. Pandey et al. \cite{Pandey2025} talked about password-less authentication mechanisms and supported the use of biometric authentication, hardware authentication tokens, mobile-based authentication, and certificate-based authentication. We already talked about the danger of biometric authentication. All the other authentication mechanisms bind users to specific devices and create huge dependencies that, in our opinion, damage flexibility. Also, there are chances of credential stealing that the authors did not talk about. Abduhari et al. \cite{Abduhari2025} also talked in support of the mechanisms that we have already discussed several times. Ahmad et al. \cite{Ahmed2025} also talked about the user behavior-based analysis that we talked about earlier. And like that, this method of theirs is also confined to a limited scope of human behavior. Pal \cite{Pal2022} also suggested the implementation of rate-limiting, breach monitoring and alerting, and blocklisting old and similar passwords. Yet, we all know that these mechanisms have problems with earlier explanations. Like Pal \cite{Pal2019}, Islam \cite{Islam2025} also suggested those techniques, along with some other extra techniques that we have covered already. Thomas et al. \cite{Thomas2019} also took the path of leveraging leaked passwords and warning users of weak or breached passwords, and suggesting that users choose stronger passwords.\\

On the other hand, our proposed framework presents a balanced mechanism where a good amount of user involvement has been ensured while keeping in mind that it doesn't bring too much burden on a user. A user can choose the policy, customize it, and switch from one rule to another at will. User memorability with this mechanism will surely be increased because we tried to design this mechanism as a game that brings joy while users play with it, and that will obviously increase the relationship between the user and the security implementation. Here, the system has a lot of involvement in this mechanism, which makes important decisions, and the difference between other mechanisms and our mechanism is that we have not solely relied on the system to do all the work. From users’ significant involvement to systems' substantial effort, our ultimate goal is to develop a system that is “perfectly balanced,” “as all things should be.”\\

We have reviewed a total of 42 papers that proposed defenses against several attacks on passwords. In the figure above, we have shown the frequencies of defense proposals by various authors that will help to visualize which mechanisms authors are supporting the most. Among the 42 papers, the recommendation of complex passwords has appeared the highest 10 times, that can be seen in Figure 12.
\begin{figure}[htbp]
\centering
\includegraphics[width=0.5\textwidth]{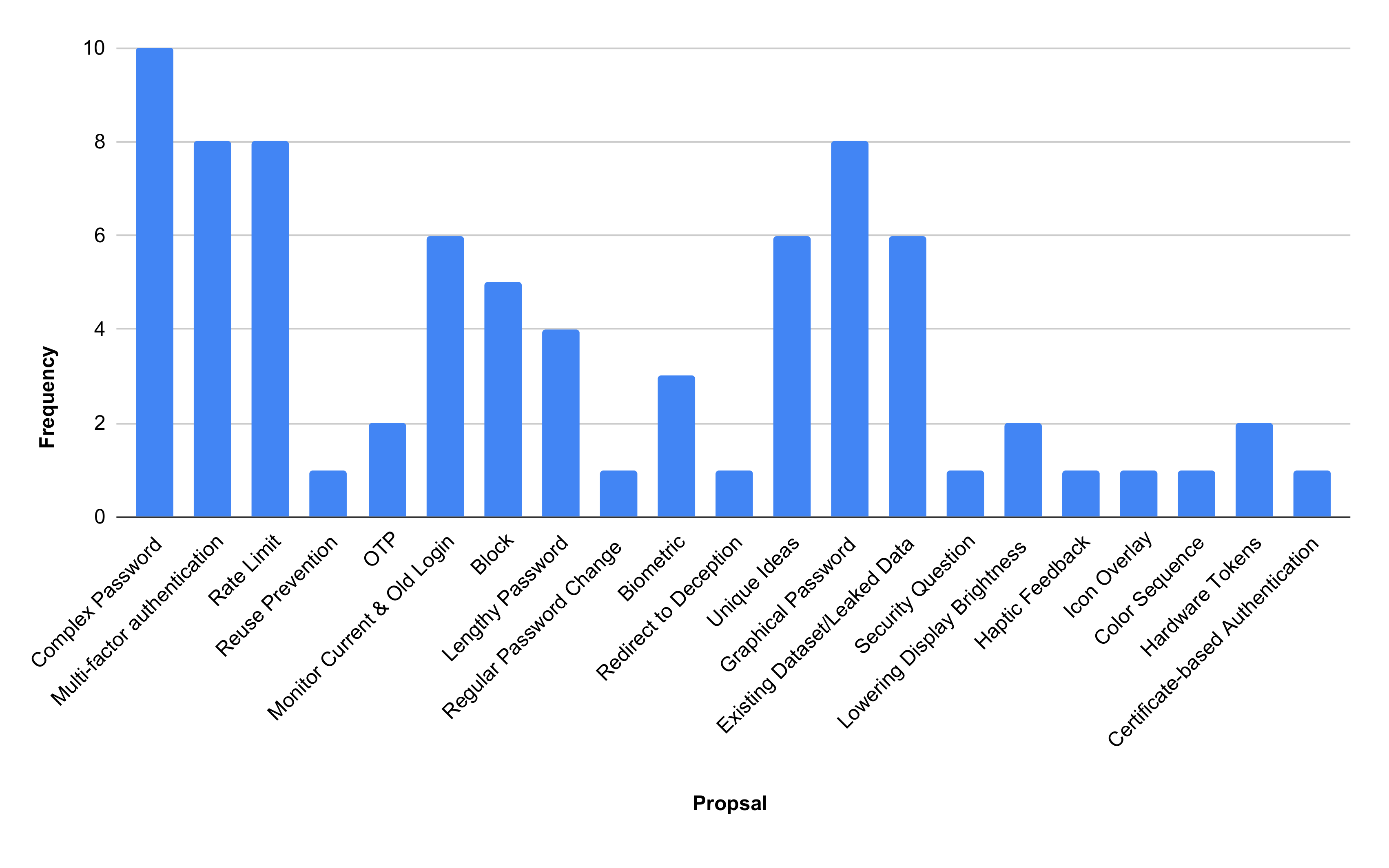}
\caption{Delineation of the frequencies of proposals from previous studies by other authors.}
\label{fig12}
\end{figure}
\vspace{-0.5em}

\begin{table}[htbp]
\caption{Frequency percentage of proposals from previous studies by other authors.}
\label{tab5}
\centering
\begin{tabular}{@{}lc@{}}
\toprule
\textbf{Proposed Solutions} & \textbf{Percentage(\%)} \\
\midrule
Complex Password & 23.80 \\
Multi-factor authentication & 19.04 \\
Rate Limit & 19.04 \\
Reuse Prevention & 2.38 \\
OTP & 4.76 \\
Monitor Current \& Old Login Data & 14.28 \\
Block & 11.90 \\
Lengthy Password & 9.52 \\
Regular Password Change & 2.38 \\
Biometric & 7.14 \\
Redirect to Deception & 2.38 \\
Unique Ideas & 14.28 \\
Graphical Password & 19.04 \\
Existing Dataset/Leaked Data & 14.28 \\
Security Question & 2.38 \\
Lowering Display Brightness & 4.76 \\
Haptic Feedback & 2.38 \\
Icon Overlay & 2.38 \\
Color Sequence & 2.38 \\
Hardware Tokens & 4.76 \\
Certificate-Based Authentication & 2.38 \\
\bottomrule
\end{tabular}
\end{table}

In Table 5, we have depicted the percentage of occurrence of proposed solutions of previous authors in the field of password security. There, we can see suggestions of setting complex passwords appear most of the time. Also, multi-factor authentication, rate limit, and graphical password setting showed a dominating existence in the list of authors’ choices. There were some unique ideas proposed by the authors, but they have their own limitations that we have discussed above already. 

The difference between other mechanisms and this mechanism is that an adversary can target a mechanism and make a fool of it with different clever measures, or the system owner implements defense for one type of threat while others are left unnoticed, whereas our mechanism is a rendezvous of multiple techniques, which provide multi-angle defenses. We have already shown the complete architecture of our framework and explained what it will be able to do and what kind of potential it holds to provide solutions to different sectors. Due to the dynamic nature, the Password Spraying attack can also be prevented with this mechanism.

We found through our study so far that some authors proposed unique methods that, if implemented, would require the whole system that is now in operation to be ditched. On the other hand, our system complies with the current implementations with minimal modification.

Also, for the dynamic password mechanism in which passwords will change automatically based on users’ decisions, as we have not discussed under which circumstances the password will be changed, we, for now, can say that the participants will require new passwords based on their preferences on every successful login during their follow-up meetings.

\section{Conclusion}
This paper proposed AdaptAuth, a multi-layered authentication framework that integrates the Password Dissection Mechanism with the Dynamic Password Policy Mechanism and enriches them with 173 behavioral and credential-based features. The framework addresses the long-standing trade-off between usability and security by combining user-driven password adaptability with system-driven behavioral profiling.

The contributions of this work include an improved password dissection method, the introduction of the Time Rule for dynamic passwords, and a comprehensive feature set for machine learning–based decision-making. These mechanisms collectively strengthen resistance against brute force, credential stuffing, and other guessing-based attacks while reducing the likelihood of false account lockouts.

Although the present work remains at a conceptual stage, it establishes a foundation for a scalable and adaptive authentication ecosystem. Future work will focus on empirical validation, optimization for real-world deployment, and the incorporation of privacy-preserving techniques. Overall, the proposed framework demonstrates strong potential for advancing password-based authentication toward a more secure and user-centric model. 

\section{Limitations}
Though this work contributes a robust mechanism that is unique and comes with a promise of thwarting attempts to log in by people with ill intentions, it has some limitations, too.
\begin{enumerate}
\item \textbf{Storage overhead:} This mechanism demands more storage than current password security practices need. There will be multiple hashes that obviously will increase the amount of data, and it has yet to be tested. Though we are hoping that, for better security, this limitation could be ignored.
\item \textbf{Computational time:} We have already said that this mechanism will produce multiple hashes, and that will take more time than what is needed in current practice. However, with the advancement of technologies, we expect that this will not be a problem for the powerful machines of the current generation to handle.
\end{enumerate}

\section{Future Works}
\begin{enumerate}
\item Our immediate next work in this project will be doing a few surveys to train the machine and find human-computer relationships. These surveys will span months or years based on the needs of the study, and as we said earlier, university students are good candidates for it because we can get them for a long time.
\item A key consideration for deployment is defining the circumstances for applying dynamic passwords without negatively impacting the user experience. User tolerance for frequent authentication is a well-known usability constraint. Therefore, calibrating the policy's application to maximize security while maintaining user acceptance is a targeted objective for subsequent empirical research.
\item Until now, our system has focused only on authentication mechanisms. If, in the future, social media giants, digital marketplaces and other important platforms share information beyond authentication about one’s behavior like strange purchasing patterns (e.g., high-value items only), rapid follow-up actions accessing sensitive features (password reset, settings, change email, disable MFA), session starting times, etc. immediately after login then this mechanism could have more accuracy in finding out adversaries.
\item Our ultimate goal is to develop a centralized authentication method for all the online platforms that require account creation. If it becomes possible, then uniquely identifying a user will be more efficient, and the burden of handling security will be significantly reduced for the digital platforms.
\end{enumerate}

\bibliographystyle{IEEEtran}
\bibliography{references}

\onecolumn
\section*{APPENDIX A}
\begin{longtable}{>{\raggedright\arraybackslash}p{0.15in}>{\raggedright\arraybackslash}p{5in}>{\raggedright\arraybackslash}p{0.8in}}
  \toprule
  \textbf{No.} & \textbf{Features} & \textbf{Data Type} \\
  \midrule
  \endfirsthead
  \toprule
  \textbf{No.} & \textbf{Features} & \textbf{Data Type} \\
  \midrule
  \endhead
  \bottomrule
  \endfoot
  1. & Browser type/version (e.g., Chrome 123.0) & String \\
  2. & Operating system and version (e.g., Windows 11, Android 14) & String \\
  3. & Device type (e.g., desktop, mobile, tablet) & String \\
  4. & Device time & Time format \\
  5. & Installed fonts or plugins (where available) & String \\
  6. & Screen resolution and color depth & String \\
  7. & Touch vs. keyboard input capabilities & Integer/Boolean \\
  8. & User-Agent string & String \\
  9. & Canvas fingerprinting hash (HTML5 feature for subtle device uniqueness) & String \\
  10. & AudioContext Fingerprinting & Class data \\
  11. & Multiple accounts accessed from the same IP in a short time & Boolean \\
  12. & Same fingerprint across many IPs or accounts & Boolean \\
  13. & Missing browser entropy (no screen size, no plugins, etc.) & Class data \\
  14. & IP-based country, region, city & Class data \\
  15. & ISP/Organization & String \\
  16. & Latitude and longitude (approximate) & String \\
  17. & Geolocation velocity (distance and time from last known location) & Class data \\
  18. & Region familiarity score (based on previous successful logins) & Integer/Floating point \\
  19. & Timezone and system clock offset & String \\
  20. & IP address reputation (blacklisted, clean, dynamic/static) & String/Integer \\
  21. & Is VPN detected? (Yes/No) & Boolean \\
  22. & Is Proxy detected? (Yes/No) & Boolean \\
  23. & Is TOR exit node? (Yes/No) & Boolean \\
  24. & Connection type (wired, mobile data, public Wi-Fi) & String \\
  25. & ASN (Autonomous System Number) -- can help trace institutional access & String \\
  26. & Time of login (HH:MM) & Time format \\
  27. & Day of the week & String \\
  28. & Mean successful login session starting time window from historical data (e.g., usually between 8--9 PM) & Time format \\
  29. & Failed login interval variance (compared to normal rhythm) & String \\
  30. & Device/browser familiarity & Boolean \\
  31. & Availability of cookie/token from previous session & Boolean \\
  32. & Number of successful logins from current device & Integer \\
  33. & First-seen timestamp of device & Time format \\
  34. & Changes in system locale or keyboard language settings & Boolean \\
  35. & Login attempt frequency in last X minutes/hours & Integer \\
  36. & Number of failed logins in a session & Integer \\
  37. & Login from multiple ips & Boolean \\
  38. & Number of login from multiple ips, if any & Integer \\
  39. & Login attempt from unknown ip(s) & Boolean \\
  40. & Total number of successful logins & Integer \\
  41. & Total number of failed logins & Integer \\
  42. & Total number of failed login from known ip(s) & Integer \\
  43. & Elapsed time from initiation of login attempt until successful authentication & Floating Point \\
  44. & Mouse movement during login & Class Data \\
  45. & Touch data from a mobile device during login & Class data \\
  46. & Scrolling speed on a particular page & Floating point \\
  47. & Window-focus events (e.g., switching tabs before login) & Class data \\
  48. & Clipboard access detection (pasting passwords vs typing) & Boolean \\
  49. & Touch event heatmap (for mobile --- helps in distinguishing automation/bots) & Class data \\
  50. & Click behavior against buttons, textboxes & Class data \\
  51. & Click pattern against a particular page & String \\
  52. & Key press and release timings & Time format \\
  53. & Dwell time (duration key is pressed) & Floating point \\
  54. & Flight time (interval between keys) & Floating point \\
  55. & Order of positions of mistakes & Integer \\
  56. & Typing speed for a full password for every failed login attempt & Floating point \\
  57. & Typing speed for a full password for every successful login attempt & Floating point \\
  58. & Shift key long pressed or short pressed & String \\
  59. & Caps Lock button used & Boolean \\
  60. & TAB button pressed to switch between textboxes & Boolean \\
  61. & Special character and Number switch button & Boolean \\
  62. & Length of password during every login button pressed (temporary data) & Integer \\
  63. & Length of password same/bigger/smaller & String \\
  64. & Number of times password length mismatch & Integer \\
  65. & Number of times length exceeded original password value & Integer \\
  66. & Number of times length fell short of the actual password length & Integer \\
  67. & Incident of appearances of the same length of passwords & Boolean \\
  68. & Number of times the same length of passwords appeared & Integer \\
  69. & Positions of mistakes & Integer \\
  70. & Number of positions of mistake(s) in every login attempt & Integer \\
  71. & Ambient character or not & Boolean \\
  72. & Character case alteration & Boolean \\
  73. & Error frequency for a particular position in a session & Integer \\
  74. & Total error frequency for a particular position for all time & Integer \\
  75. & Number of times character case alteration in a position for all time & Integer \\
  76. & Number of times character case alteration until a single login button press & Integer \\
  77. & Number of times character case alteration in a session & Integer \\
  78. & Number of times ambient values are entered in a position for all time & Integer \\
  79. & Number of times ambient values are entered for all positions combined for all time & Integer \\
  80. & Number of times ambient values are entered until a single login button pressed & Integer \\
  81. & Number of times ambient values are entered in a session & Integer \\
  82. & Number of times wrong special character input in a position for all time & Integer \\
  83. & Number of times wrong special character input for all the positions combined for all time & Integer \\
  84. & Number of times wrong special character input until a single login button is pressed & Integer \\
  85. & Number of times wrong special character input in a session & Integer \\
  86. & A user uses single or multi-class values & String \\
  87. & Number of value classes appeared in the current login session(temporary data) & Integer \\
  88. & Number of positions based on multiple value classes' appearance (temporary data) & Class data \\
  89. & Identification of correct values amid heterogeneous inputs at a position & Boolean \\
  90. & Correct input is single or multiple in a position & String \\
  91. & Multiple positions had correct input & Boolean \\
  92. & Single correct input is the only one that is entered in the very first & Boolean \\
  93. & Single correct input is random & Boolean \\
  94. & Number of wrong tries before the correct input appears & Integer \\
  95. & Number of times of having correct input & Integer \\
  96. & Number of positions of having correct input & Integer \\
  97. & Failed login contains correct password(s) but the sequence is wrong & Boolean \\
  98. & Same class character(s)(temporary data) & Boolean \\
  99. & User input distant value or not & Boolean \\
  100. & Distant value's distant character(s) entered & Boolean \\
  101. & Total number of distant value inputs in a position in a session & Integer \\
  102. & Total number of distant value inputs in a position for all time & Integer \\
  103. & Distant value's ambient character(s) entered & Boolean \\
  104. & Number of ambient values of a distant value has been used & Integer \\
  105. & Positions where distant values were entered until a single login button is pressed & Integer \\
  106. & Positions where distant values were entered in a session & Integer \\
  107. & Positions where distant values were entered for all time & Integer \\
  108. & Total position numbers where distant values have been entered until a single login button is pressed & Integer \\
  109. & Total position numbers where distant values have been entered in a session & Integer \\
  110. & Total number of distant value inputs in all positions combined in a session & Integer \\
  111. & Total number of distant value inputs in all positions combined for all time & Integer \\
  112. & Distance level of the tried characters (close, far) & String \\
  113. & Matching percentage increased/decreased/remained unchanged because of distant value input & String \\
  114. & Keys pressed in a login session & String \\
  115. & Sequence of key pressing & String \\
  116. & Password pasting & Boolean \\
  117. & Matching percentage increased/decreased/remained unchanged because of ambient key input & String \\
  118. & Ambient value led to login success & Boolean \\
  119. & Distant value led to login success & Boolean \\
  120. & Rule name & String \\
  121. & User's frequent mistakes & String \\
  122. & Frequency of rule changes & Integer \\
  123. & Deviated from the rule & Boolean \\
  124. & Number of deviations from the rule in one session & Integer \\
  125. & Number of deviations from the rule for all time & Integer \\
  126. & Rule repetition threshold (e.g., user rotates rules every 3 logins) & String \\
  127. & Decoy rule existence (can be used for the security challenge) & Boolean \\
  128. & Decoy position altered (high red flag) & Boolean \\
  129. & Position(s) chosen for rule application & Integer \\
  130. & Position(s) where decoy rule applied & Integer \\
  131. & Matching percentage & Floating point \\
  132. & Position(s) of mismatched values & Integer \\
  133. & Error increased/decreased/unchanged & String \\
  134. & The percentage of error is unchanged with the new positional problem arising and the old one getting fixed & Boolean \\
  135. & Position that got fixed & Integer \\
  136. & Position that got a new error & Integer \\
  137. & Number of attempts before solving a positional error & Integer \\
  138. & Number of failed attempts before a successful login & Integer \\
  139. & CAPTCHA solving speed based on CAPTCHA type for a single user & Floating point \\
  140. & Average CAPTCHA solving speed based on CAPTCHA type for all users & Floating point \\
  141. & Types of CAPTCHAs a user has tried & String \\
  142. & CAPTCHA solving accuracy based on individual CAPTCHA for a single user & Floating point \\
  143. & CAPTCHA solving accuracy based on individual CAPTCHA for all users & Floating point \\
  144. & Session-based CAPTCHA solving accuracy & Floating point \\
  145. & Overall CAPTCHA solving success rate by a user & Floating point \\
  146. & Average CAPTCHA solving success rate by all users for an individual CAPTCHA & Floating point \\
  147. & CAPTCHA complexity classification based on a single user & String \\
  148. & CAPTCHA complexity classification based on all user & String \\
  149. & Dwell time for CAPTCHA image & Floating point \\
  150. & Flight time for CAPTCHA image & Floating point \\
  151. & Time to solve CAPTCHAs in a session & Floating point \\
  152. & Time from the appearance of CAPTCHA to start solving & Floating point \\
  153. & The user uses the backspace button & Boolean \\
  154. & Number of times the backspace button was used in a complete password & Integer \\
  155. & Number of times backspace button used in a session for each user & Integer \\
  156. & User empties textbox & Boolean \\
  157. & User removed last typed character & Boolean \\
  158. & User removed character in the middle & Boolean \\
  159. & Values removed by one backspace button press at a time or long press & String \\
  160. & Dwell time for the backspace button & Floating point \\
  161. & Positions the user used the backspace button & Integer \\
  162. & User switch rule & Boolean \\
  163. & Rule chosen during account creation & String \\
  164. & Rule embracement rate for a particular rule & Floating point \\
  165. & Rule leaving rate for a particular rule & Floating point \\
  166. & Dwell time on a rule that was set during the account creation & Floating point \\
  167. & Time between new rule acceptance and leaving & Floating point \\
  168. & Number of times a rule has been chosen by users & Integer \\
  169. & Number of login attempts failed against every rule, session-wise & Integer \\
  170. & Number of login attempts failed against every rule for all time & Integer \\
  171. & Return to the previous rule & Boolean \\
  172. & Rule false positive & Boolean \\
  173. & User shifted from dynamic rule to static rule & Boolean \\
  \bottomrule
\end{longtable}
\end{document}